\documentclass{jfm}

\usepackage{graphicx}
\usepackage{natbib}
\usepackage{color}
\usepackage{lineno}

\begin{document}

\newtheorem{lemma}{Lemma}
\newtheorem{corollary}{Corollary}
\renewcommand{\vec}[1]{\mbox{\boldmath$ #1 $}}

\shorttitle{Stability of dancing {\it Volvox} } 
\shortauthor{T.~Ishikawa, T.~J.~Pedley, K.~Drescher \& R.~E.~Goldstein} 

\title{Stability of dancing \emph{Volvox}}

\author
 {
Takuji Ishikawa\aff{1},
T.~J.~Pedley\aff{2}
  \corresp{\email{t.j.pedley@damtp.cam.ac.uk}},
K.~Drescher\aff{3}
  \and 
  Raymond~E.~Goldstein\aff{2}
  }

\affiliation
{
\aff{1}
Department of Finemechanics, Tohoku University, 6-6-01, Aoba, Aramaki, Aoba-ku, Sendai 980-8579, Japan
\aff{2}
Department of Applied Mathematics and Theoretical Physics, University of Cambridge, Centre for Mathematical Sciences, Wilberforce Road, Cambridge CB3 0WA, UK
\aff{3}
Max Planck Institute for Terrestrial Microbiology, Philipps-Universit\"{a}t Marburg, Karl-von-Frisch-Strasse 10, D-35043 Marburg, Germany
}

\maketitle

\begin{abstract}
Biflagellate algal cells of the genus \emph{Volvox} form spherical colonies that propel themselves, vertically upwards in still fluid, by the coordinated beating of thousands of flagella, that also cause the colonies to rotate about their vertical axes. When they are swimming in a chamber of finite depth, pairs (or more) of \emph{Volvox carteri} colonies were observed by Drescher et al. [\textit{Phys. Rev. Lett.} {\bf 102}, 168101 (2009)] to exhibit hydrodynamic bound states when they are close to a rigid horizontal boundary. When the boundary is above, the colonies are attracted to each other and orbit around each other in a `waltz'; when the boundary is below they perform more complex `minuet' motions. These dances are simulated in the present paper, using a novel `spherical squirmer' model of a colony in which, instead of a time-independent but $\theta$-dependent tangential velocity being imposed on the spherical surface (radius $a$; $\theta$ is the polar angle), a time-independent and uniform tangential shear stress is applied to the fluid on a sphere of radius $(1+\epsilon)a, \epsilon \ll 1$, where $\epsilon a$ represents the length of the flagella. The fluid must satisfy the no-slip condition on the sphere at radius $a$. In addition to the shear stress, the motions depend on two dimensionless parameters that describe the effect of gravity on a colony: $F_g$, proportional to the ratio of the sedimentation speed of a non-swimming colony to its swimming speed, and $G_{bh}$, that represents the fact that colonies are bottom-heavy. $G_{bh}$ is the ratio of the time scale to swim a distance equal to the radius, to the time scale for gravitational reorientation of the colony's axis to the vertical when it is disturbed. In addition to reproducing both of the dancing modes, the simulations are able to determine values of $F_g$ and $G_{bh}$ for which they are stable (or not); there is reasonable quantitative agreement with the experiments.

\end{abstract}


\section{Introduction}\label{sec:1}

\emph{Volvox} is a genus of algae, several species of which form spherical, free-swimming colonies consisting of up to 50,000  somatic cells embedded in an extracellular matrix on the surface, with interior germ cells that develop into a small number of colonies of the next generation. The colony has an anterior-posterior axis of symmetry and each somatic cell bears a pair of beating flagella that enable the colony to swim approximately parallel to this axis. Each cell's flagella beat in approximately the same direction (relative to the colony), i.e. in a plane that is offset from a purely meridional plane by an angle of $10^\circ - 20^\circ$. This offset causes the colony to rotate about its axis as it swims \citep{Hoops1993}; the rotation is always clockwise when viewed from its anterior pole. In still water colonies tend to swim vertically upwards, on average, because they are bottom-heavy (daughter colonies being clustered towards the rear), although they are slightly (about 0.3\%) denser than water and therefore sediment downwards when the flagella are inactivated. The beating of the flagella of cells at different polar angles, $\theta$, has been observed, in colonies held stationary on a micro-pipette, to be coordinated in the form of a symplectic metachronal wave, which propagates from anterior to posterior in the same direction as the power stroke of the flagellar beat \citep{Brumley:2012}. Modelling suggests that hydrodynamic interactions between the flagella of different cells, coupled with flagellar flexibility, provide the mechanism for the coordination \citep{Niedermayer2008,Brumley:2012,Brumley:2015MW}. A detailed survey of the physics and fluid dynamics of green algae such as \emph{Volvox} has been given by \cite{Goldstein:arfm}.

The radius $a$ of a \emph{Volvox} colony increases with age (the lifetime of a \emph{V. carteri} colony is about 48 hours) though the number and size of cells do not. \cite{Drescher:2009vn} measured the free-swimming properties of many colonies of \emph{V. carteri} of different radii. Results for the mean upswimming speed $W$, sedimentation speed $V_g$, angular velocity $\Omega$, mean density difference between a colony and the surrounding fluid (inferred from $V_g$) and timescale $\tau$ for reorientation by gravity when the axis is disturbed from the vertical (a balance between viscous and gravitational torques) are shown in Fig.~1(a-e). Note that, if the colony were neutrally buoyant, the swimming speed would be $U = W + V_g$, and that the colony Reynolds number is always less than about $0.15$ so that the hydrodynamics is dominated by viscous forces. Note too that the largest colonies cannot make upwards progress; they naturally sink towards the bottom of the swimming chamber, even when their flagella continue to perform normal upswimming motions.

\cite{Drescher:2009vn} also observed the behaviour of \emph{V. carteri} colonies as they swim up towards a horizontal glass plane above or sink towards a horizontal plane below. In the former case, the flagella on a colony that is close to the upper surface continue beating and applying tangential thrust to the nearby fluid. Since the fluid is prevented from flowing from above, the flagellar beating pulls in fluid horizontally from all round and thrusts it downwards. This was observed by seeding the fluid with 0.5 $\mu$m polystyrene microspheres and the velocity field measured in horizontal and vertical planes using Particle Image Velocimetry (PIV) \citep{Drescher:2009vn}.

The simplest model of a swimming colony \citep{Short:2006} ascribes the total mean force exerted by the flagella to a uniform tangential shear stress exerted on the spherical surface, with components $f_\theta$ and $f_\phi$ in the directions of the polar angle $\theta$ and the azimuthal angle $\phi$. \cite{Drescher:2009vn} estimated $f_\theta$ and $f_\phi$, as functions of colony radius $a$, from the measured values of $U$, $V$ and $\Omega$ and low-Reynolds-number hydrodynamics (the Stokes Law and the equivalent for rotation):
 \begin{equation}\label{eq:1.1}
f_\theta = 6\mu (W+V_g)/\pi a, ~~ f_\phi = 8\mu \Omega/\pi,
\end{equation}
where $\mu$ is the fluid viscosity. The estimated values corresponded to a few pN per flagellar pair, as also found by \cite{Solari:etal2006}. It can be inferred from these results that there is a critical colony radius, $a_c$, at which a colony far from any boundaries will hover at rest. For the experiments shown in Fig.~1, $a_c \approx 300\mu$m.
 

When two \emph{Volvox} colonies of approximately the same size, with $a < a_c$, were introduced into the chamber, and when they were both spinning near the upper surface, they were observed to attract each other and to orbit around each other in a bound state, termed a 'waltz', by \cite{Drescher:2009vn} (Fig.~2a and movie M1). When the individual angular velocity $\Omega$ was about 1 rad s$^{-1}$, the orbiting frequency $\omega$ was about 0.1 rad s$^{-1}$. The mutual attraction is consistent with the radial inflow of small particles to a single colony, and the rate of approach of two nearby colonies close to the top wall can be well approximated by treating each colony as a point Stokeslet at the sphere's centre, together with its image system in the plane \citep{Drescher:2009vn}.  These results provided the first quantitative
experimental verification of the prediction by \cite{Squires2001} of a wall-mediated attraction between downward-pointing Stokeslets near an upper no-slip surface.

However, the orbiting is not a direct consequence of colony 2 translating in the swirling velocity field generated by colony 1, for example, because an isolated colony does not generate a swirl velocity field. The overall torque on a colony is zero; therefore the azimuthal ($\phi$-direction) torque generated by the beating flagella is balanced by an equal and opposite viscous torque on the colony as a whole, as if the flagella were trying to crawl along the inside surface of a shell of fluid, but succeeds only in pushing the spherical colony surface in the opposite direction. The orbiting could come about because of near-field effects as the colonies approach each other: the rotation rate of colony 1 is reduced by viscous forces in the gap between the two colonies.   To predict the rate of orbiting, \cite{Drescher:2009vn} added a vertical rotlet at the centre of each sphere, together with its image in the plane and, assuming that the surface of each sphere was rigid, used lubrication theory to calculate the force and torque exerted by one sphere on the other for a given rotation rate $\Omega$. The torque provides the rotlet strength and this, together with the force, determines the orbiting frequency,
\begin{equation}\label{eq:1.2}
\omega \approx 0.069 ~ \log { (d/2a)}\Omega,
\end{equation}
where $d$ is the separation between the two spherical surfaces.  Equation (1.2) is close to the average of measurements on 60 different waltzing pairs.

Some pairs of colonies with $a\approx{a_c}$, which individually hover, form time-dependent bound states near the bottom of the chamber, with one colony above the other, both colonies oscillating horizontally back and forth. This motion was called a `minuet' by \cite{Drescher:2009vn}. In this regime the state of perfectly aligned colony axes is unstable, the flow generated by the swimming of one colony tilting and moving the other one away, while the latter's bottom-heaviness and swimming tend to bring it back (see Fig.~2(b) and movie M2). The distance between two minuetting colonies is large enough for lubrication effects to be negligible, so \cite{Drescher2010thesis} modelled each one as a vertical gravitational Stokeslet, the resulting sedimentation being balanced by steady swimming with speed $U$, directed at a small angle $\theta^{(m)} (m=1,2)$ to the vertical. This angle is determined by a gyrotactic balance between viscous and gravitational torques, the latter arising from bottom-heaviness. The height of each colony above the chamber bottom was taken to be fixed.

Thus the model consisted of two vertical Stokeslets located at the centres of the spheres, ${\bf x}^{(m)}$, plus their image systems in the horizontal plane below (\cite{Blake:image}; see Fig. 3). The motion of sphere $m$ is given by

\begin{equation}\label{eq:1.3}
\frac{d{\bf x}^{(m)}}{dt} = {\bf u}^{(m)} + U {\bf p}^{(m)},
\end{equation}
where ${\bf u}^{(m)}$ is the velocity field at sphere $m$ generated by the Stokeslet of the other sphere and its image system in the plane, and 
\begin{equation}\label{eq:1.4}
{\bf p}^{(m)} = (\sin{ \theta^{(m)}},0,\cos{ \theta^{(m)}})
\end{equation}
is the unit orientation vector of sphere $m$ (note that axes have been taken such that ${\bf p}^{(m)}$ lies in the $x_1 x_3$ plane, $x_3$ being vertically upwards - see Fig.~3). If we consider $m=1$, take the Stokeslet strength of each sphere to be $(0,0,-8\pi\mu F)$, and consider only the $x_1$-component of (1.3), then the results of \cite{Blake:image} give

\begin{equation}\label{eq:1.5}
u^{(1)}_1 = -F\left[\frac{r_1 r_3}{r^3} - \frac{R_1 R_3}{R^3} -2H\frac{\partial}{\partial R_3}\left(\frac{HR_1}{R^3} - \frac{R_1 R_3}{R^3}\right)\right],
\end{equation}
where  $H$ is the height of the centre of sphere 2 above the plane, assumed constant. Here ${\bf r} = {\bf x}^{(1)} - {\bf x}^{(2)} = (r_1,0,r_3)$ and ${\bf R} = {\bf x}^{(1)} - {\bf x}^{(2')} = (r_1,0,r_3 +2H)$, where ${\bf x}^{(2')}$ is the image of ${\bf x}^{(2)}$; $r$ and $R$ are the magnitudes of $\bf{r}$ and $\bf{R}$, respectively. For constant height $r_3 = h$, small displacement $r_1$, and small angles $\theta^{(m)}$, equations (1.5) and (1.3) with $m=1$ reduce to

\begin{equation}\label{eq:1.6}
\frac{dx^{(1)}_1}{dt} = -\frac{Fr_1}{h^2} (1-\beta_1) + U \theta^{(1)},
\end{equation}
where
\begin{equation}\label{eq:1.7}
\beta_1 = \frac{h^2(h^2+8hH+6H^2)}{(h+2H)^4}.
\end{equation}
The corresponding expression for $\frac{dx^{(2)}_1}{dt}$ is also obtained from (1.5) by replacing $[r_1, H, h, \theta^{(1)}]$ by $[-r_1, H+h, -h, \theta^{(2)}]$, which leads to 
\begin{equation}\label{eq:1.8}
\frac{dx^{(2)}_1}{dt} = -\frac{Fr_1}{h^2} (1+\beta_2) + U \theta^{(2)},
\end{equation}
where
\begin{equation}\label{eq:1.9}
\beta_2 = \frac{h^2(-h^2+4hH+6H^2)}{(h+2H)^4}.
\end{equation}
In addition,
\begin{equation}\label{eq:1.10}
\frac{d{\bf p}^{(1)}}{dt} = \frac{1}{B}\left\{\left[{\bf k} - ({\bf k}\cdot {\bf p}^{(1)})\right]{\bf p}^{(1)} + \frac{1}{2}{\bf \omega}^{(1)}\wedge {\bf p}^{(1)}\right\},
\end{equation}
where $\bf{k}$ is a vertical unit vector, and ${\bf \omega}^{(1)}$ is the vorticity at ${\bf x}^{(1)}$ due to the sphere at ${\bf x}^{(2)}$ and the image system, so (1.9) becomes
\begin{equation}\label{eq:1.11}
\frac{d\theta^{(1)}}{dt} = -\frac{1}{B}\sin{ \theta^{(1)}} - \frac{F x^{(1)}_1}{r^3} \gamma \approx -\frac{\theta^{(1)}}{B} - \frac{F x^{(1)}_1}{h^3} \gamma
\end{equation}
where $\gamma = 1- h^3/(h+2H)^3$, and similarly for $d \theta^{(2)}/dt$. Here $B = 6 \mu/(l \rho g)$, where $l$ is the distance from a colony's centre of buoyancy to its centre of mass, is the timescale for gyrotactic reorientation.

If we write $\xi = x^{(1)}_1 - x^{(2)}_1$, $\Theta = \theta^{(1)} - \theta^{(2)}$ and $\beta = (\beta_1 + \beta_2)/2$, the system of linearised equations (1.6), (1.8), (1.11) reduces to 
\begin{equation}\label{eq:1.12}
\frac{d\xi}{dt} = \frac{2F}{h^3} \beta \xi +U \Theta ,~~~~~~~~\frac{d\Theta}{dt} = -\frac{1}{B} \Theta - \frac{2F}{h^3} \gamma \xi. 
\end{equation}
Assuming that $\xi$ and $\Theta$ are proportional to e$^{\lambda t}$, (1.12) gives a quadratic equation for the eigenvalues $\lambda$ whose roots are

\begin{equation}\label{eq:1.13}
\lambda = \frac{1}{2}\left[-\frac{1}{B} + \frac{2F\beta}{h^2} \pm \sqrt{\left(-\frac{1}{B} + \frac{2F\beta}{h^2}\right)^2 - \frac{8FU \gamma}{h^3} }\right].
\end{equation}
Thus the equilibrium steady state $(\xi = \Theta = 0)$ is unstable if $B > B_c = \frac{h^2}{2F\beta}$. Moreover the bifurcation at $B = B_c$ is a Hopf bifurcation if the quantity in the square root is negative. The parameter values of the experiments by \cite{Drescher:2009vn} ($B=14$s, $h=600 \mu$m, $H=450\mu$m, $F=6.75\times10^4 \mu$m$^2/$s, $U =300 \mu$m/s), which give $B_c = 13.2$s, satisfy these inequalities, which is consistent with the observed oscillations. \cite{Drescher:2009vn} computed the solution to the nonlinear system (1.3) and (1.10) and indeed found that it exhibited limit-cycle oscillations for these parameters over a limited range of values of $B$: 12s $< B <$ 20s. It should be noted that the nonlinear version of this model is still only a coarse approximation, as it neglects vertical motions of the two colonies, as well as their rotation about the vertical, which can give rise to orbiting motions when the colony axes are not vertical.

The purpose of this paper is to provide a more detailed fluid mechanical understanding of the pairwise interactions of \emph{Volvox} by means of an improved model of the above phenomena, which confirms and extends the modelling results of \cite{Drescher:2009vn}. We will simulate the flow due to two identical, spinning squirmers in a semi-infinite fluid with a rigid horizontal plane either above or below, for a range of realistic values of the relevant parameters. In section 2 the problem is specified precisely and the numerical method (using the Boundary Element Method, or BEM) described. The results are presented in section 3 for the waltz and section 4 for the minuet. They will consist of representative movies of both the waltz and the minuet (in supplementary material) with careful comparison with the experiments of \cite{Drescher:2009vn}. In particular we seek to delineate regions of parameter space in which the dancing modes are stable and investigate what happens when they are not.



\section{Basic equations and numerical methods}\label{sec:2}

\subsection{A Volvox model}

A single colony is modelled as a steady `spherical squirmer', modified from that used previously to study the hydrodynamic interactions between two such model organisms and their behaviour in suspensions (\cite{Ishikawa:2006,Ishikawa:2007rheology,Ishikawa:2007diffusion,Ishikawa:2008coherent,Pedley2016IMA}). In those studies the velocity on the spherical surface of the squirmer was taken to be purely tangential and prescribed as a function of polar angle $\theta$, while remaining symmetric about the orientational axis, represented by unit vector $\textbf{p}$. Moreover, the azimuthal, $\phi$, component of velocity was taken to be zero. Thus an isolated squirmer of uniform density would `swim' in the direction of $\textbf{p}$, at a constant speed, $U$, but would not rotate. Here, instead of the surface velocity, we prescribe the mean shear stress ${\bf f}_s$ generated by the beating flagella of \emph{Volvox} as acting tangentially at a radius $\alpha a$, where $\alpha = 1 + \epsilon$ and $\epsilon a$ is proportional to the length of a flagellum (Fig. 4a); there is no slip on the colony surface $r = a$. The shear stress has components $f_\theta$ and $f_\phi$ in the $\theta$- and $\phi$- directions, i.e. ${\bf f}_s = (f_r, f_\theta, f_\phi)$ with $f_r = 0$. Prescribing stresses not velocities is probably more realistic, especially when colonies come close to each other or to a fixed boundary, and especially because it permits no slip on the surface $r=a$. Non-zero $f_\phi$ means that colony rotation is automatically included. The model is still greatly oversimplified because the stresses are taken to be constant, independent of both time and position (i.e. $\theta$ and $\phi$). A similar `stress and no-slip' squirmer model was used, but not fully analysed, in the computations of \cite{Ohmura:2018}. We also note that the model bears some relation to the `traction-layer' model for ciliary propulsion proposed by \cite{Keller1975}, and to the model studied by \citet{Short:2006}.

Solution of the Stokes equations shows that the swimming speed and rotation rate of a neutrally buoyant squirmer in an infinite fluid are given by
\begin{equation}\label{eq:2.1}
U = \frac{a f_\theta}{\mu} \frac{\pi}{6} \frac{4\alpha^3 - 3\alpha^2 - 1}{4\alpha} ,~~~ \Omega = -\frac{f_\phi}{\mu} \frac{\pi}{8} (\alpha^3 - 1)
\end{equation}
(see Appendix A for details). Thus, for small $\epsilon = \alpha - 1$, the dimensionless shear stresses are given by $(af_\theta / (\mu U)) = 4/(\pi \epsilon)$, which can be approximately inferred from the experimental measurements of Fig.~1(a-c), as long as a value of $\epsilon$ is assumed (this is discussed further in section 5). Moreover, the stresslet strength, which is important in determining the effect of micro-organisms on the fluid flow around them (\cite{Simha2002,Saintillan2008}), is identically zero, so according to this model \emph{Volvox carteri} is approximately a neutral squirmer \citep{MichelinLauga:2010}, consistent with the observations of \cite{Drescher:2010flowfield}.

As for previous models \citep{Ishikawa:2007diffusion,Ishikawa:2007rheology} we can incorporate bottom-heaviness by supposing that the centre of mass of the sphere is displaced from the geometric centre by the vector $-l\bf{p}$, so when $\bf{p}$ is not vertical the sphere experiences a torque $-l\bf{p}\wedge \bf{g}$ that tends to rotate it back to vertical (${\bf g} = - g{\bf k}$ is the gravitational acceleration). The relevant dimensionless quantity representing the effect of bottom-heaviness relative to that of swimming is
\begin{equation}\label{eq:2.2}
G_{bh} = \frac {4\pi \rho g a l}{3 \mu U} = \frac {8 \pi a}{B U},
\end{equation}
where $\rho$ is the density of the fluid. When $G_{bh} = 8 \pi$, the angular velocity of a neutrally buoyant colony that is oriented horizontally in an infinite fluid becomes $U/a$.
We also add a point Stokeslet at the centre of the sphere to represent the negative buoyancy of a \emph{Volvox} colony. The dimensionless quantity representing the effect of sedimentation relative to that of swimming is
\begin{equation}\label{eq:2.3}
F_g =  \frac {4\pi \Delta \rho g a^2}{3 \mu U},
\end{equation}
where $\Delta \rho$ is the density difference between a colony and the fluid. When $F_g = 6 \pi$, the sedimentation velocity in an infinite fluid is $U$.

\subsection{Basic equations}

Since the colony Reynolds number is always less than about $0.15$, we neglect inertia. In the Stokes flow regime, the velocity $\bf{u}$ is given by an integral equation over the colony surface $S_c$ and the shell of shear stress $S_f$ as \citep {Pozrikidis:book}
\begin{equation}
\label{eq:bi}
{\bf u}({\bf x}) = - \frac{1}{8\pi\mu} \int_{S_c} {\bf J}({\bf x}, {\bf y}) \cdot {\bf q}({\bf y})dS_c({\bf y})
- \frac{1}{8\pi\mu} \int_{S_f} {\bf J}({\bf x}, {\bf y}) \cdot {\bf f}_s ({\bf y})dS_f({\bf y}),
\end{equation}
where ${\bf J}$ is the Green function for a flow bounded by an infinite plane wall (\cite {Blake:image}), and $\bf{q}$ is the traction force. $\vec{q}$ is defined as $\vec{q} = \vec{\sigma}
\cdot \vec{n} = \left( -p \vec{I} + 2 \mu \vec{E} \right) \cdot \vec{n}$, where
$\vec{\sigma}$ is the stress tensor, $p$ is the
pressure and $\vec{E}$ is the rate of strain.
On the surface of the rigid sphere, the no-slip boundary condition is given by
\begin{equation}
{\bf u}({\bf x}) = {\bf U}
+ \vec{\Omega} \wedge {\bf r}
~~,~~~~{\bf r} \in S_c,
\label{bem.bc}
\end{equation}
where ${\bf U}$ and $\vec{\Omega}$ are the translational and rotational velocities of the colony.

The shear stress ${\bf f}_s$ expresses the thrust force per unit area generated by the flagellar beat. The thrust force should be balanced by the viscous drag force and the sedimentation force. Thus, the force condition for a colony can be given as
\begin{equation}
\int_{S_c} {\bf q} dS_c + \int_{S_f} {\bf f}_s dS_f
 + \frac{4 \pi a^3 \Delta \rho}{3} {\bf g} + {\bf F}_{rep}= 0 .
\end{equation}
Here ${\bf F}_{rep}$ is the non-hydrodynamic repulsive force between colonies and between a colony and a wall. Although lubrication flow can prevent a rigid sphere colliding with a plane wall, the shear stress shell can easily collide with a plane wall or another shear stress shell. In the case of a real \emph{Volvox}, the collision tends to deform the flagella, and the repulsive force may be generated by the elasticity of flagella. Here, we do not model such a complex phenomenon, but follow \cite{BradyBossis:stokesiandyn} and \cite{Ishikawa:2007rheology} and use the following function
\begin{equation}
{\bf F}_{rep} =
\alpha_1 \frac{\alpha_2 \exp(-\alpha_2 \lambda)\bf{s}}
{(1-\exp(-\alpha_2 \lambda)) s} 
\label{repulsive}
\end{equation}
where ${\bf s}$ is the centre-to-centre vector between two colonies or the normal vector from the wall to the colony centre; $\alpha_1$, $\alpha_2$ are dimensionless coefficients and $\lambda$ is the minimum separation between two  shear stress shells or between a shear stress shell and the wall, non-dimensionalized by $a$. The coefficients used in this study are $\alpha_1 = 10$ and $\alpha_2 = 10$
\textcolor[rgb]{1,0,0}{}
for colony-wall interactions, whereas $\alpha_1 = 1$ and $\alpha_2 = 10$ for colony-colony interactions. The parameters were chosen to avoid collision while keeping computational efficiency. Since the colony surfaces are at least $2 \epsilon a$ apart in the present study, the  repulsive force remains much smaller than the lubrication forces, and is much less significant than in \cite{Ishikawa:2006}, in which the gap could become infinitely small. The minimum separation obtained with these parameters is of the order of $10^{-2}a$.

The torque condition is given by
\begin{equation}
\int_{S_c} {\bf r} \wedge {\bf q} dS_c
 + \int_{S_f} {\bf r} \wedge {\bf f}_s dS_f
 - \frac{4 \pi a^3 \rho l}{3} {\bf p} \wedge {\bf g} = 0 .
\end{equation}
The repulsive force does not contribute to the torque balance.

\subsection{Numerical methods}

The governing equations are discretized by a boundary element method (BEM) \citep{Pozrikidis:book}. By combining the governing equations and the boundary condition, a set of linear algebraic equations can be generated. Each spherical surface of a colony is discretized by 320 triangles, while each spherical shear stress shell is discretized by 1280 triangles. The numerical integration is performed using 28-point Gaussian polynomials, and the singularity is solved analytically. Time-marching is performed using a fourth-order Runge-Kutta method. The details of these numerical methods can be found in \cite{Ishikawa:2006}.

The coordinate axes are taken as shown in Fig.~4b. Gravity acts in the $- {\bf e}_3$ direction, i.e. ${\bf k} = {\bf e}_3$, and an infinite plane wall exists at $e_3 = 0$. When we investigate a waltzing motion beneath the wall, colonies are placed in the negative $e_3$ half-space. When we investigate a minuet motion above the wall, on the other hand, colonies are placed in the positive $e_3$ half-space. ${\bf p}^{(m)}$ is the orientation vector of colony $m$. The angle of ${\bf p}^{(m)}$ from ${\bf e}_3$ is defined as $\theta_{p}^{(m)}$.

Parameter values are varied so as to cover experimental conditions. By assuming that the relaxation time, $B$, defined as $6 \mu / \rho g l$, is about 14 seconds \citep{Drescher:2009vn} and the colony swims about one body length per second, $G_{bh}$ is about 2. In the present study, $G_{bh}$ is varied in the range $0 - 100$. Small and young \emph{Volvox} swim faster than the sedimentation speed, though large and old \emph{Volvox} cannot swim upwards. In order to cover both conditions, $F_g$ is varied in the range $0 - 9 \pi$.
The tilt angle of the flagellar beating plane with respect to the colonial axis was about 15$^\circ$  \citep{Drescher:2009vn}. We thus set $\arctan(f_{\phi}/f_{\theta}) =$ 15$^\circ$ throughout this study; $\epsilon$ is set as $0.05$ on the basis of experimental observations \citep{Brumley:2012}.


\section{Waltzing motion beneath a top wall}\label{sec:3}

We first calculate the flow field around a single colony hovering beneath a plane wall. The colony is directed vertically upwards, and the hovering motion is stable when the colony is sufficiently bottom-heavy. The results for velocity vectors in the $e_1 - e_3$ plane are shown in Fig. 5a ($G_{bh}=25$ and $F_g = 3 \pi$). We see that fluid is pulled in radially towards the colony and then goes downward. The white broken arrows in the figure schematically show the vortex structure. The computed flow field is similar to that observed experimentally.
When two colonies hover beneath a wall, they are attracted to each other due to the inward suction. The change with time of the centre-to-centre distance $s$ between two colonies with $G_{bh}=25$ and $F_g = 3 \pi$ is shown in Fig. 5b, where $t_0$ is the time of collision. The broken line indicates experimental results from \cite{Drescher:2009vn} averaged over 60 colonies. The solid line indicates the simulation result, in which the time scale is dimensionalized by using the characteristic time of $a/U =
0.5$ sec. The attraction velocity increases as the distance decreases, which is captured in the simulation.

Two nearby colonies beneath a wall orbit around each other in a `waltz', as stated above. Fig. 6 and Supplementary Movie 3 show the waltzing motion reproduced by the simulation under the condition of $G_{bh} = 25$ and $F_g = 3 \pi$.  We see that two colonies orbit around each other with a constant rotation rate. The radius of the orbiting is approximately
1.07, so the two shear stress surfaces are very close to contact.

In order to discuss the stability of the waltzing motion, we calculated the change of orientations and distance between two nearby colonies. Fig. 7a shows the definitions of parameters used in the analysis.
Let ${\bf x}^{(m)} = ( x_1^{(m)}, x_2^{(m)}, x_3^{(m)})$ and ${\bf p}^{(m)} = ( p_1^{(m)}, p_2^{(m)}, p_3^{(m)})$ respectively be the position vector and the orientation vector of colony $m$.  For simplicity, we assume that the two colonies align in the $e_2$-direction, \emph{i.e.} $x_1^{(1)} = x_1^{(2)}$ and $x_3^{(1)} = x_3^{(2)}$. The orientation vectors were set as $p_1^{(1)} = -p_1^{(2)},  p_2^{(1)} = -p_2^{(2)}$ and $p_3^{(1)} = p_3^{(2)}$, so that a rotation of $\pi$ around the $e_3$-axis leaves the configuration unchanged.

The length of the centre-to-centre vector is set as $2.14a$.  The colour-coded values of $ds/dt$ indicate the separation velocity between the two colonies, i.e. $ds/dt < 0$ is attractive, whereas $ds/dt > 0$ is repelling. $\phi_p$ is the angle of the projection vector of ${\bf p}$ in the $e_1 e_2$ plane from the line connecting the two colony centres. Because of the condition $p_1^{(1)} = -p_1^{(2)},  p_2^{(1)} = -p_2^{(2)}$, $\phi_p$ is the same for each colony; $\theta_p$, defined in Fig. 4b, is also the same for each colony.

Fig. 7b shows the results of the stability analysis with $G_{bh} = 25$ ($F_g = 3 \pi$), in which stable waltzing motion was observed. The horizontal axis indicates $\phi_p$, the vertical axis indicates $\theta_p$, and the components of the white vectors are $d \phi_p / dt$ and $d \theta_p / dt$ at given $\phi_p$ and $\theta_p$. Moreover, the colour indicates the separation velocity $ds/dt$. By following the white vectors and considering the separation velocity, we can understand how the configurations of two colonies change with time. The black dot in Fig. 7b indicates the stable point, where a point sink of the white vector field exists with $ds / dt \le 0$. We can conclude that the waltzing motion with $G_{bh} = 25$ is stable with respect to small fluctuations in the colony configurations.

In the case of $G_{bh} = 5$ ($F_g = 3 \pi$), on the other hand, there is no stable point (Fig. 7c). Thus, colonies with $G_{bh} = 5$ eventually repel each other and do not show the stable waltzing motion. Fig. 8 shows the phase diagram for the stability of waltzing motion in $G_{bh} - F_g$ space. The waltzing becomes unstable in the bottom grey region, while it is stable in the top white region. The boundary lies between $G_{bh} = 5$ and 10, and $F_g$ has little influence on it.
The mean $G_{bh}$ value in the experiments can be estimated as about 1.8 \citep{Drescher:2009vn}, which is smaller than the stable limit in the simulation. There might be two possibilities to explain the discrepancy. First, the flagella beat might be disturbed in the experiments due to interaction with the top glass wall. If the flagella beat is disturbed, the torque generated by the flagella will be reduced, which effectively increases their bottom-heaviness to stabilize the vertical orientation of the colony. Another possibility is that it was only colonies with large $G_{bh}$ that were observed in the experiment, because only they could stay near a top wall for a sufficiently long time.

Next, we discuss the mechanism of the waltzing motion. For simplicity, we again assume the simple configuration shown in Fig. 7a. The coordinate system, forces, torques and velocities are defined in Fig. 9. In the Stokes flow regime, the motions of two rigid spheres in the presence of a plane wall can be described by using the mobility tensor \citep{KimKarrila:book}. Hence, the orbiting velocity of colony 1, $U_1^{(1)}$, which is equivalent to the orbiting rotation rate multiplied by the orbiting radius, can be given as follows
\begin{equation}
\label{eq:mobility}
U_{x1} = M_{1,1} F_1^{(1)} + M_{1,5} T_2^{(1)} + M_{1,6} T_3^{(1)}
       + M_{1,7} F_1^{(2)} + M_{1,11} T_2^{(2)} + M_{1,12} T_3^{(2)} ,
\end{equation}
where $M_{i,j}$ is the $(i,j)$ component of the $12 \times 12$ mobility tensor. $F_2, F_3$ and $T_1$ do not contribute to $U_1$ due to the symmetry of the problem. $M_{i,j}$ can be calculated by BEM in the stable waltzing configurations with $G_{bh}=25$ and $F_g = 3 \pi$, and the results are $(M_{1,1}, M_{1,5}, M_{1,6}, M_{1,7}, M_{1,11}, M_{1,12}) = 10^{-2} (2.39, -0.13, -0.8, 0.31, -0.01, -0.50)$. The forces and torques can also be calculated by BEM by fixing two colonies in space with the active shear stress ${\bf f}_s$. The results are $(F_1^{(1)}, T_2^{(1)}, T_3^{(1)}, F_1^{(2)}, T_2^{(2)}, T_3^{(2)}) = (-3.1, 1.5, -13.9, 3.1, -1.5, -13.9)$. The largest positive contribution comes from $M_{1,12} T_3^{(2)} = 0.069$, and other major positive contributions are $M_{1,7} F_1^{(2)} = 0.010$ and $M_{1,6} T_3^{(1)} = 0.011$. The largest negative contribution comes from $M_{1,1} F_1^{(1)} = -0.074$. Thus, one may roughly say that the orbiting velocity $U_1^{(1)}$ is mainly generated by $T_3^{(2)}$ and inhibited by $F_1^{(1)}$. $T_3^{(2)}$ is induced on the colony as a reaction torque from the flagellar beat. Negative $F_1^{(1)}$ is induced because the traction force ${\bf q}^{(1)}$ acting in regions A and A$^\prime$ in Fig. 9 are different. In region A, ${\bf q}^{(1)}$ is induced by the shear stress of colony 1, ${\bf f}_{s}^{(1)}$. In region A$^\prime$, on the other hand, ${\bf q}^{(1)}$ is induced by the shear stress of both colonies, ${\bf f}_{s}^{(1)}$ and ${\bf f}_{s}^{(2)}$, which tend to cancel each other out. Thus, smaller ${\bf q}^{(1)}$ is generated in region A$^\prime$ than in A.

The angular velocity of an individual spinning with $G_{bh} = 25$ and $F_g = 3 \pi$ is $\Omega_3^{(1)} \approx -0.41$. The angular velocity of orbiting, $\omega$ ($= -2U_1^{(1)}/s$) is about 0.013. The ratio of angular velocity of orbiting to that of spinning is about 0.03 in the simulation, which is considerably smaller than the experimental value of 0.19 from \cite{Drescher:2009vn}. The ratio, however, can be modified dramatically by reducing the value of $G_{bh}$, as shown in Fig. 10. When $G_{bh}$ becomes small, the colony orientations tend to tilt from the $e_3$-axis and appear to follow each other. Such inclination dramatically reduces $F_1^{(1)}$ in Eq.(\ref{eq:mobility}), and therefore increases $\omega$. We see from Fig. 10 that the effect of $G_{bh}$ on $\omega$ is significant, though that of $F_g$ is small.


\section{Minuet motion above a bottom wall}\label{sec:4}

When colonies become large as they age, so that $F_g$ exceeds approximately $6\pi$, the sedimentation velocity exceeds the swimming velocity. Such colonies stay near a bottom wall and sometimes interact with each other, as discussed in section 1. Before going into the details of two-colony interactions, we first calculate the flow field around a solitary colony. Fig. 11a shows the simulated velocity vectors around a colony with $F_g = 9 \pi$ hovering stably at a height of approximately 3.2 (non-dimensionalised with $a$) over a bottom wall ($G_{bh} = 5$). The wall is at $x_3 = 0$, and the $x_3$-axis is taken as shown in the figure. The colony is directed vertically upwards. We see that strong downward flow is generated around the colony. A toroidal vortex, shown by white arrows, is observed at the side of the colony. The height of stable hovering decreases as $F_g$ increases, as shown in Fig. 11c. However, even for $F_g$ as large as $9\pi$, a colony exhibits a positive upswimming velocity when its height above the wall is less than the height of stable hovering (Fig. 11b).

Next we examine the `minuetting' bound state of two colonies near a bottom wall. We show three examples; in each case colony 1 has $F_g = 7.5 \pi$ and colony 2 has $F_g = 9 \pi$. Both colonies are assumed to have the same $G_{bh}$ value, and $G_{bh}$ is varied from 2 to 6. Other parameters of the colonies, such as $a$ and $\epsilon$, are the same. Colonies 1 and 2 are initially placed at (-1.5, 0, 5) and (1.5, 0, 3), respectively. Trajectories of the two colonies near the bottom wall for time $t$ in the range $0-100$ are shown in Fig. 12. When $G_{bh} = 2$ (cf. Fig. 12a and Supplementary Movie 4), the two colonies attract each other when they are apart, but repel each other when they are close to contact. Attraction and repulsion are repeated, forming the `minuet' bound state. In order to discuss the oscillation of trajectories in the horizontal direction, we calculate the distance between the two colonies projected onto the $e_1 e_2$ plane. The results are shown in Fig. 13. We see that the horizontal distance oscillates with amplitude up to 3.5 in the case of $G_{bh} = 2$.

In the case of $G_{bh} = 3$ (cf. Fig. 12b and Supplementary Movie 5), the minuet motion is still observed, but the amplitude of the oscillation in the horizontal distance decreases to about 2 (cf. Fig. 13). This is because the orientation change induced by hydrodynamic interactions is suppressed by the bottom-heaviness. 
We see that the centres of two colonies form almost two-dimensional trajectories up to $t = 30$, though the trajectories become gradually 3-dimensional and the two colonies eventually orbit around each other in a bound state. We note that the direction of orbiting relative to the direction of spin, in this case, is opposite to the `waltzing motion' observed near a top wall. Moreover, it seems that two-dimensional minuet motion can be unstable in the direction perpendicular to the plane. In the case of $G_{bh} = 6$, the two colonies eventually align vertically (cf. Fig. 12c and Supplementary Movie 6). Similar alignment was observed in the experiment (\cite{Drescher:2009vn} and Supplementary Movie 2). The horizontal distance, shown in Fig. 13, gradually converges to zero in this case.
We note that even when two colonies have the same $F_g$ values, such as $F_{g,1} = F_{g,2} = 7.5 \pi$ or $9 \pi$, we observed minuet motion, orbiting around each other or vertical alignment depending on the $G_{bh}$ values and the initial positions.

In Fig. 14, we show the phase diagram of two-colony interactions near a bottom wall ($F_g = 7.5 \pi$ and $9 \pi$). The black circle in the figure indicates unstable motion, in which the centre-to-centre distance between the two colonies exceeds $10a$. The white circles indicate the minuet motion or orbiting around each other in a bound state. The black triangles indicate vertical alignment, in which the distance in the $e_1 - e_2$ plane is less than $0.3a$ for $t = 90-100$. We see that the colonies show the minuet motion when $G_{bh}$ is in the appropriate range, while they align vertically when $G_{bh}$ is large. The effects of the $G_{bh}$ values of colonies 1 and 2 on the stability are almost symmetric.

Last, we compare the present numerical results with the theory of \cite{Drescher:2009vn} in which two-colony interactions were analyzed by assuming far-field hydrodynamics (see section 1 above). Each colony was assumed to have the same Stokeslet strength, i.e. the same $F_g$, but different equlibrium heights above the bottom boundary. This is not fully compatible with our results in Fig. 11, where different equilibrium heights follow from different values of $F_g$. In our simulations we assume that colony 1 has $F_g = 6.5 \pi$ and colony 2 has $F_g = 9 \pi$, so that the two colonies have very different heights of stable hovering (cf. Fig. 11c) and may interact mainly in the far-field. Colonies 1 and 2 are initially placed at (-1.5, 0, 7) and (1.5, 0, 3), respectively. Trajectories of the two colonies near the bottom wall for time $t$ in the range $0-1000$ or until centre-to-centre distance exceeds $10a$ are shown in Fig. 15. When $G_{bh} = 0.1$ (cf. Fig. 15a), the two colonies first show minuet motion, but eventually move
apart from each other. The centre-to-centre distance between the two colonies, in this case, is shown in Fig. 15d. We see that the distance oscillates due to the minuet motion, but gradually increases with time. In the range $t > 200$, the distance is larger than 4, so near-field hydrodynamics does not play a major role. Hence, we may say that the minuet motion in this case is unstable even in the far-field. In Eq. (\ref{eq:1.12}), the $G_{bh}$ value for unstable interactions can be estimated as $G_{bh} < 0.017$ by assuming $H = h = 2a$ as in Fig. 15c. The present results illustrate that the minuet motion can be unstable even with $G_{bh} = 0.1$. 

\textcolor[rgb]{1,0,0}{}
When $G_{bh} = 0.3$ (cf. Fig. 15b), the two colonies first show minuet motion, then move apart from each other at around $t = 500$ (cf. Fig. 15d), then come close once again at around $t = 730$, and eventually separate fully. The second separation  is induced by the near-field interactions at around $t = 730$, so the minuet motion becomes unstable due to the near-field hydrodynamics in this case. When $G_{bh} = 1$ (cf. Fig. 15c), the two colonies show a stable hydrodynamic bound state, in which they first show almost two-dimensional minuet motion, before the trajectories become gradually three-dimensional due to the instability of the two-dimensional minuet motion in the direction perpendicular to the plane. At around $t=600$, the two colonies orbit  each other, and the motion continues until $t=1000$. These results illustrate that near-field hydrodynamics also plays an important role in the hydrodynamic bound states of squirmers.

\section{Discussion}\label{sec:5}

\textcolor[rgb]{0,0,1}{}The boundary-element computations in this paper, using the `shear-stress and no-slip' spherical squirmer model for a swimming micro-organism, have succeeded in simulating the dancing motions performed by colonies of \emph{Volvox carteri} in the experiments of \cite{Drescher:2009vn}. 

In the case of the waltzing bound state of pairs of \emph{Volvox} colonies near the top wall of the chamber, the computations confirm the approximate analysis of \cite{Drescher:2009vn}, based on the earlier work of 
\cite{Squires2001}, which utilizes point Stokeslets and rotlets at the centres of the two colonies, and lubrication theory for the space between them when they are close together. In addition we examined the stability of the waltzing state, and found that it is stable if the bottom-heaviness parameter $G_{bh}$ exceeds a critical value between 5 and 10, more or less independently of the gravitational Stokeslet parameter $F_g$. A typical experimental value of $G_{bh}$ was estimated from the data in Fig.~1 to be around 1.8, which is below the critical value although the `waltzing' appeared stable; the reason for this discrepancy has not been firmly established, though it seems likely that (a) the flagellar beating is reduced in the narrow gap between the colonies and the plane wall above, due to mechano-sensing and a reduction of the flagellar beat frequency, or due to flagella sticking to the glass, as has been observed for \emph{Tetrahymena} \citep{Ohmura:2018}, and (b) only colonies with larger values of $G_{bh}$ would stay near the top surface for long enough to attract a neighbour into the waltz.

For the minuet bound state near the bottom boundary,  \cite{Drescher:2009vn} gave calculated results for (in our notation) $a = 300 \mu m, H/a = 2, h/a = 1.5$ and $F_g = 0.75$. Their Fig. 5c shows stable vertical alignment of two colonies for $B < 12$s ($G_{bh} > 2.1$) and limit cycle oscillations for $12$s $< B < 20$s ($2.1 > G_{bh} > 1.3$, becoming unstable for larger $B$ (smaller $G_{bh}$). Our simulations have the same qualitative features, but the threshold values of $G_{bh}$ are significantly smaller. As discussed in section 4, this discrepancy could be a consequence of the assumption of the same value of $F_g$ for both colonies, but in addition it may also have one of the following two causes: (a) the far-field assumption may not be accurate enough when the distance between the colonies is less than $10a$, and (b) the minuet motion can become three-dimensional and the heights of the two colonies vary with time, though in Eq.(\ref{eq:1.12}) two-dimensional trajectories with constant heights were assumed.

Finally, it is appropriate to give further discussion to the `shear-stress and no-slip' squirmer model itself; here we neglect the density difference between the sphere and the fluid, so $F_g = 0$ and sedimentation is absent. The formulae (2.1) relating the mean swimming speed $U$ and the mean angular velocity $\Omega$ to the shear stresses $f_\theta$ and $f_\phi$ exerted at $r=(1+\epsilon)a$ give

\begin{equation}
\label{eq:5.1}
f_\theta = \frac{4}{\pi} \frac{\mu U}{\epsilon a} ~~\mathrm{and} ~~f_\phi = \frac{8}{3\pi} \frac{\mu \omega}{\epsilon},
\end{equation}
to leading order in $\epsilon$ for $\epsilon \ll 1$. These are not the same as given by \cite{Drescher:2009vn}, quoted in (1.1) above. Our model recognises that the shear stress is effectively exerted by the beating flagella, at their tips in the power stroke, and lower down in the recovery stroke. The model requires that the resultant velocity field satisfies the no-slip condition on the (rigid) spherical surface of the \emph{Volvox} colony, as well as the zero-Stokeslet condition for a self-propelled body. The earlier model balanced the total force exerted by the shear stress against the viscous (Stokes)  drag on an inert sphere pulled through the fluid at the same speed. This ignores the fact that the force on the rigid sphere at $r=a$ consists of both the hydrodynamic (shear stress and pressure) force and the equal and opposite reaction force experienced by the flagella and transmitted by them to the rigid sphere. The force exerted by the flagella not only drives the outer flow, but also the high-shear flow in the flagella layer. Put another way, the previous model balanced 
the rate of viscous energy dissipation in the flow in $r > a_0$ driven by the shear stress against the rate of working of the Stokes drag on the inert sphere, but ignored the energy dissipation between the shear stress shell and the no-slip spherical boundary, i.e. in the flagella layer. If we model the flow in this layer as a uniform shear flow, as in Fig.16, the total rate of energy dissipation in the layer is $D_1 = \mu\times 4\pi a^2 h \times (f_\theta /\mu)^2$, where $h=\epsilon a$, which scales as $\epsilon a^3 f_\theta ^2 /\mu$. The dissipation in the outer flow, assumed to scale similarly to that for a translating rigid sphere, i.e. $6\pi \mu a U^2$, which from (5.1) scales as $D_2 \sim \epsilon^2 a^3 f_\theta^2 /\mu$, is formally an order of magnitude smaller than that in the layer. In \emph{V. carteri} $\epsilon$ is between $0.05$ and $0.1$ \citep{Solari:etal2011}, which is not very small, so in view of numerical factors we cannot say that $D_1 \gg D_2$ but we can be confident that the dissipation rate in the layer is at least as important as that outside it. It follows that a greater shear stress is required to achieve the same swimming speed than in the previous model.
Similar considerations apply to the angular velocity $\Omega$ and the zero-torque condition.

The consequence of the new formulae (5.1) is that the shear stresses for a sphere with $a=200\mu$m, $U=380\mu$m/s and $\Omega=1.3 $rad/s (Fig. 1), and flagella length $\epsilon a=15\mu$m, are $f_\theta \approx 1.9\times 10^{-2}$ N/m$^2$, $f_\phi \approx 1.6\times 10^{-2}$ N/m$^2$. Noting that 1N/m$^2 = 10^3$fN$/\mu$m$^2$, we see that these values are nearly a factor of 2 greater than the corresponding quantities in Fig.~1(\textit{f}); this is mainly a consequence of the additional energy dissipation and the $\epsilon ^{-1}$ factors in (5.1).

\vspace{10mm}
\begin{flushleft}
{\large {\bf Acknowledgments}}
\end{flushleft}

T.I. is supported by the Japan Society for the Promotion of Science Grant-in-Aid for Scientific Research (JSPS KAKENHI Grant No. 17H00853 \& 17KK0080). R.E.G. acknowledges support from the Gordon and Betty Moore Foundation (Grant No. 7523).

\vspace{5mm}

\appendix
\section{}
\label{Appendix:A}

Here we derive equations (2.1), with reference to Fig.~16. There is a no-slip spherical boundary at $r=a$ and uniform tangential stresses $f_\theta$ and $f_\phi$ are applied to the fluid at radius $a_0$. The squirmer is taken to swim at speed $U$ in the $\theta=0$ direction so, relative to the sphere, the velocity at infinity is $-U$ (in the $\theta=\pi$ direction). The sphere rotates with angular velocity $\Omega$ about the axis of symmetry; there is no azimuthal velocity at infinity. The squirmer swims freely, so the force and torque exerted on it by the fluid are zero.  We solve the axisymmetric Stokes equations separately for the radial and meridional velocity components, and for the swirl velocity component.

We consider the flow in two regions, $a<r<a_0$ (region $1$) and $a_0<r<\infty$ (region$2$), and represent it by stream functions $\psi^{(i)}(r,\theta), i=1,2$. In region $1$, the solution of the Stokes equation can be written

\begin{eqnarray}
\label{eq:streamfunction 1}
\psi^{(1)} &=& {\sum_{n=1}^{\infty}} \frac{1}{2} \sin{\theta} V_{n}(\theta) 
\\
&~& ~~~~ \times
\left[A_n ^{(1)} \left(\frac{a_0}{r}\right)^{n-2} + B_n^{(1)} \left(\frac{a_0}{r}\right)^{n} +  C_n^{(1)} \left(\frac{r}{a_0}\right)^{n+1} + D_n^{(1)} \left(\frac{r}{a_0}\right)^{n+3}\right],
\nonumber
\end{eqnarray}
where $V_{n}(\theta) = \frac{2}{n(n+1)} \sin{\theta} P_{n}' (\cos{\theta})$, the $P_n$ being Legendre polynomials, and $A_n^{(1)}, B_n^{(1)}, C_n^{(1)}, D_n^{(1)}$ are constants to be determined. In region $2$ the stream function is
\begin{eqnarray}
\label{eq:streamfunction 2}
\psi^{(2)} &=& -\frac{1}{2}U r^2 \sin^2 (\theta) + {\sum_{n=1}^{\infty}} \frac{1}{2} \sin{\theta} V_{n}(\theta) \\
&~& ~~~~ \times
\left[A_n ^{(2)} \left(\frac{a_0}{r}\right)^{n-2} + B_n^{(2)} \left(\frac{a_0}{r}\right)^{n} +  C_n^{(2)} \left(\frac{r}{a_0}\right)^{n+1} + D_n^{(2)} \left(\frac{r}{a_0}\right)^{n+3}\right] .
\nonumber
\end{eqnarray}
The first term incorporates the (unknown) uniform stream at infinity, and $C_n^{(2)} = D_n^{(2)} = 0$ for all $n$ so the corresponding contributions to the velocity tend to zero at infinity. Moreover, $A_1^{(2)}$ is also zero, because this is the Stokeslet term, proportional to the net force on the sphere, which is zero.

The velocity components, pressure and tangential shear stress in region $1$ are
\begin{eqnarray}
\label{eq:u_r}
u_r &=& -U\cos{\theta} + {\sum_{n=1}^{\infty}} \frac{1}{a^2} P_n(\cos {\theta}) \\
&~& ~~~~ \times
\left[A_n^{(1)}\left(\frac{a_0}{r}\right)^{n} + B_n^{(1)} \left(\frac{a_0}{r}\right)^{n+2} +  C_n^{(1)} \left(\frac{r}{a_0}\right)^{n-1} + D_n^{(1)} \left(\frac{r}{a_0}\right)^{n+1}\right] ,
\nonumber\\
\label{eq:u_theta}
u_\theta &=& +U\sin{\theta} + {\sum_{n=1}^{\infty}} \frac{1}{2a^2} V_n(\theta) 
\left[(n-2)A_n^{(1)}\left(\frac{a_0}{r}\right)^{n} + nB_n^{(1)} \left(\frac{a_0}{r}\right)^{n+2} \right. \\
&~& ~~~~~~~~~~~~~~~~~~~~~~~~~~~~~
\left. - (n+1)C_n^{(1)} \left(\frac{r}{a_0}\right)^{n-1} -(n+3) D_n^{(1)} \left(\frac{r}{a_0}\right)^{n+1}\right] ,
\nonumber \\
\label{eq:p}
p &=& \frac{2\mu}{a^4} {\sum_{n=1}^{\infty}}P_n(\cos{\theta}) 
\left[A_n^{(1)}\frac{2n-1}{n+1}\left(\frac{a_0}{r}\right)^{n+1} + D_n^{(1)} \frac{2n+3}{n}\left(\frac{r}{a_0}\right)^{n+3} \right] , \\
\label{eq:sigma}
\sigma_{r\theta} &=& - \frac{\mu}{a^4} {\sum_{n=1}^{\infty}} V_n(\theta)
\left[A_n^{(1)}(n^2 - 1)\left(\frac{a_0}{r}\right)^{n+1} +  B_n^{(1)} n(n+2)\left(\frac{a_0}{r}\right)^{n+2} \right. \\
&~& ~~~~~~~~~~~~~~~~~~~~~~~~~~~~~
\left. + C_n^{(1)} (n^2 - 1)\left(\frac{r}{a_0}\right)^{n-2} + D_n^{(1)} n(n+2)\left(\frac{r}{a_0}\right)^{n+1}\right],
\nonumber
\end{eqnarray}
with corresponding equations for region $2$. The boundary conditions at $r = a$ are $u_r = u_\theta = 0$ and at $r = a_0$ are continuity of $ u_r, u_\theta$ and the normal stress $-p + 2\mu \partial {u_r}/\partial {r}$, and the jump in $\sigma_{r\theta}$ from $1$ to $2$ is $f_\theta$. The constant $f_\theta$ can also be expanded in a series of the $V_n$:
\begin{equation}
\label{eq:sigma jump}
f_\theta = f_\theta {\sum_{n=1}^{\infty}} F_n V_n(\theta),
\end{equation}
where 
\begin{equation}
\label{eq:F_n}
F_{2l} = 0,~~  F_{2l + 1} =   \frac {(4l+3) \Gamma (l + \frac{1}{2}) \Gamma (l + \frac{3}{2})}{4 \Gamma (l+1) \Gamma(l + 2)}.
\end{equation}
Now, the object of this analysis is to calculate $U$, which appears only in the $\cos{\theta}$ and $\sin{\theta}$ terms in the above equations. Hence we need to use only the $n=1$ terms in the equations; for example the relevant contribution to $f_\theta$ is $F_1 = 3\pi/8$. A simple calculation gives the result

\begin{equation}
\label{eq:U}
U = \frac{a f_{\theta}\pi}{\mu} \frac{(4\alpha^3 - 3\alpha^2 - 1)}{24\alpha},
\end{equation}
where $\alpha = a_0/a$.

It remains to perform a similar analysis for the swirl velocity $u_\phi$. The solution of the azimuthal component of the Stokes equation in region $1$ is

\begin{equation}
\label{eq:u_phi}
u_\phi = a_0 {\sum_{n=1}^{\infty}} V_{n}(\theta) \left[G_n ^{(1)} \left(\frac{a_0}{r}\right)^{n+1} + H_n^{(1)} \left(\frac{r}{a_0}\right)^{n}\right],
\end{equation}
with corresponding azimuthal shear stress

\begin{equation}
\label{eq:az shear}
\sigma_{r\phi} = \mu r \frac{\partial (u_\phi^{(1)}/r)}{\partial{r}} = \frac {\mu r}{a_0} {\sum_{n=1}^{\infty}} V_{n}(\theta) \left[- (n+2) G_n ^{(1)} \left(\frac{a_0}{r}\right)^{n+2} + (n-1) H_n^{(1)} \left(\frac{r}{a_0}\right)^{n-1}\right].
\end{equation}
Similar equations apply to region $2$, except the the $H_n^{(2)}$ terms are all zero because the swirl velocity must tend to zero at infinity. Moreover, the torque on the body is proportional to $G_1 ^{(1)}$, so this too must be zero. The boundary conditions are that $u_\phi$ is $a \sin{\theta} \Omega$ at $r = a$ and continuous at $r = a_0$, while the jump in azimuthal shear stress at $r = a_0$ is $f_\phi F_1$. Hence we deduce that
\begin{equation}
\label{Omega}
\Omega = - \frac {f_\phi}{\mu} \frac {\pi}{8}(\alpha ^3 - 1).
\end{equation}
This completes the derivation of equations (2.1).

\bibliographystyle{jfm}
\textit{}\bibliography{tjpbibl}

\clearpage

\begin{figure}
\begin{center}
~~

\vspace{5cm}
\includegraphics[scale=0.5]{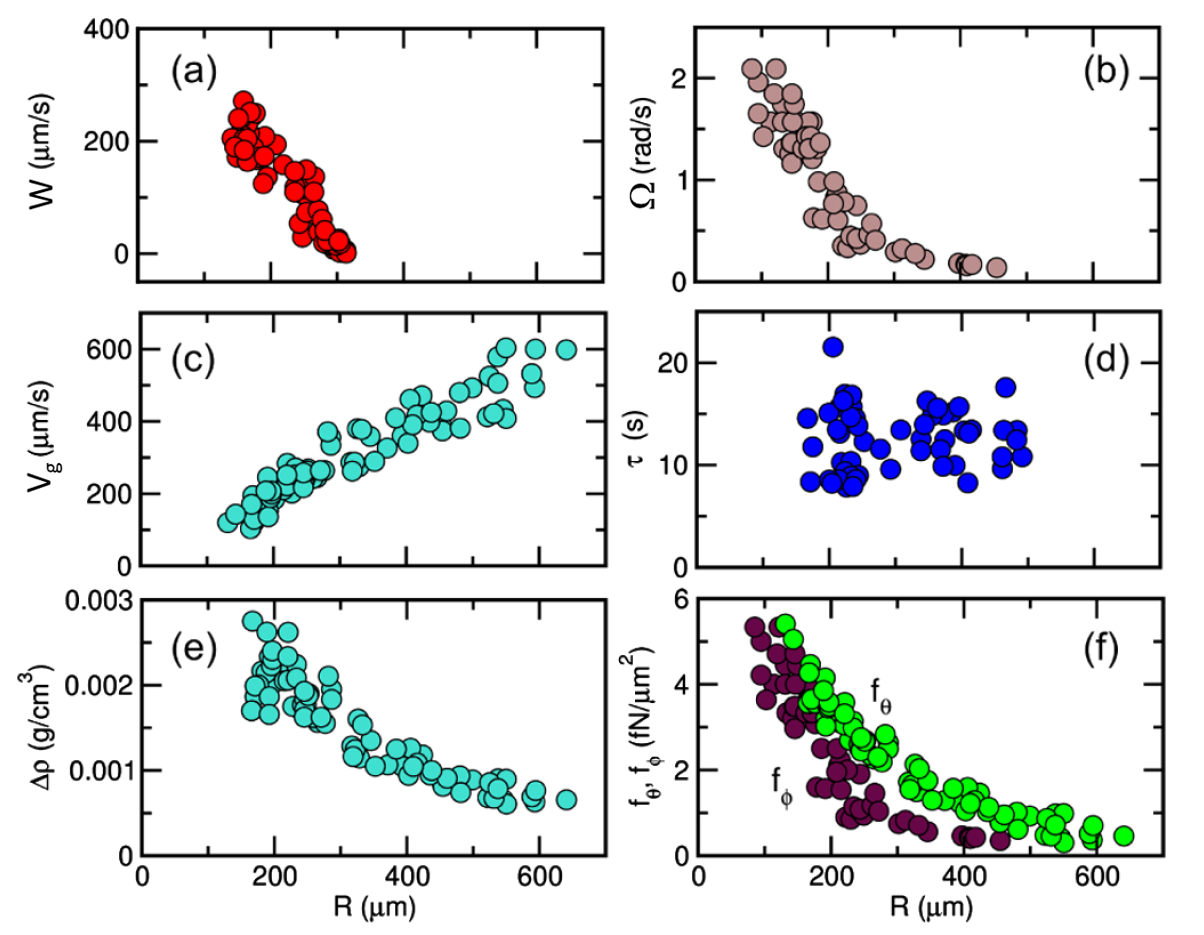}
\caption{Swimming properties of \emph{V. carteri} as a function of radius. (a) upswimming speed, (b) rotational frequency, (c) sedimentation speed, (d) bottom-heaviness reorientation time, (e) density offset, and (f) components of average flagellar force density. (From \cite{Drescher:2009vn}, Fig. 3, with permission.)
}
\label{fig1}
\end{center}
\end{figure}

\clearpage

\begin{figure}
\begin{center}
~~

\vspace{2cm}
\includegraphics[scale=0.4]{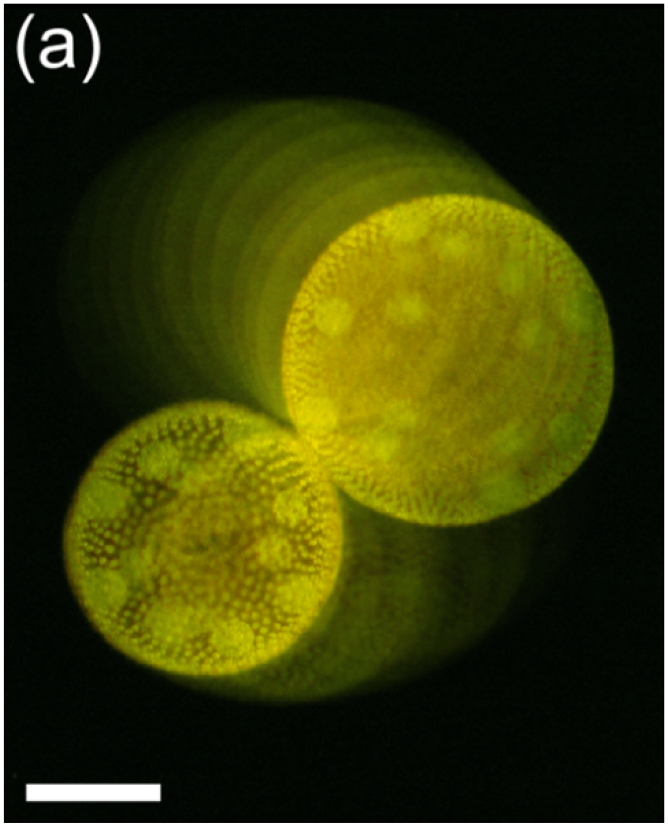}

\vspace{1cm}
\includegraphics[scale=0.35]{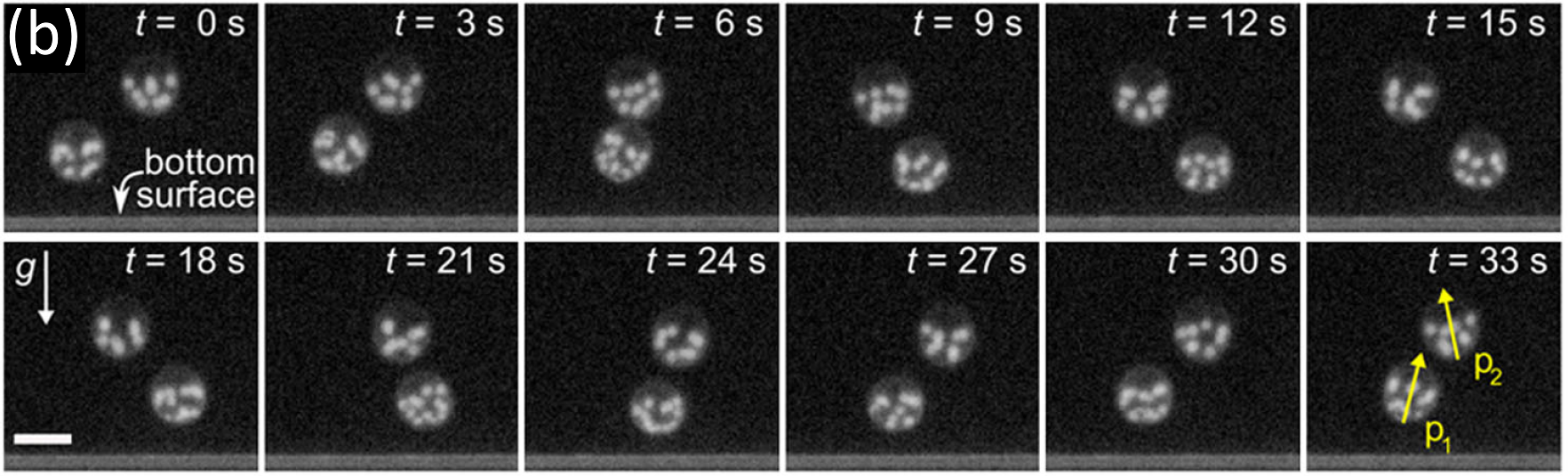}
\caption{(a) Waltzing of \emph{V.carteri}: top view. Superimposed images taken 4s apart, graded in intensity. Scale bar is 1 mm;  (b) 'minuet' bound state: side views 3s apart of two colonies near the chamber bottom. Arrows indicate the anterior-posterior axes ${\bf p}_m$ at angles $\theta_m$ to the vertical. Scale bar is 600 $\mu$m. (From \cite{Drescher:2009vn}, Fig. 1(a) and Fig. 5(a), with permission.) 
}
\label{fig2}
\end{center}
\end{figure}

\clearpage

\begin{figure}
\begin{center}
~~

\vspace{2cm}
\includegraphics[scale=0.4]{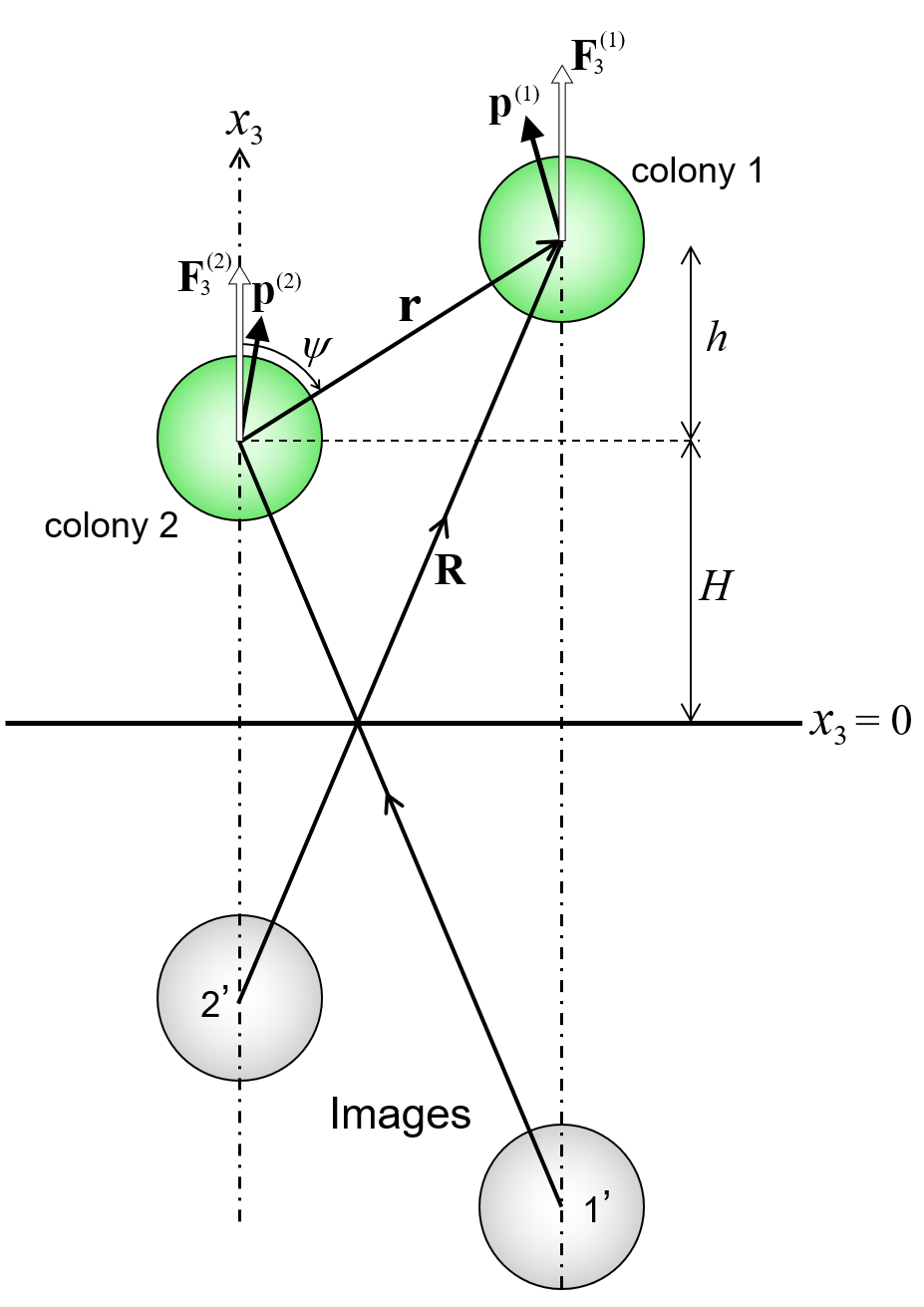}
\caption{Model for the minuet bound state: the centres of the two colonies $\bf{1}$ and $\bf{2}$ are at ${\bf x}^{(1)}$ and ${\bf x}^{(2)}$, with their images in the plane $e_3 = 0$ at ${\bf x}^{(1')}$ and ${\bf x}^{(2')}$; ${\bf r} = {\bf x}^{(1)} - {\bf x}^{(2)} , {\bf R} = {\bf x}^{(1)} - {\bf x}^{(2')}$. In the model analysed by \cite{Drescher2010thesis}, the angle  $\theta^{(m)}$ between the orientation vector of colony $m$ and the vertical is taken to be small, as is the angle $\psi$ between ${\bf r}$ and the vertical. 
}
\label{fig3}
\end{center}
\end{figure}

\clearpage

\begin{figure}
\begin{center}
~~

\vspace{3cm}
\includegraphics[scale=0.25]{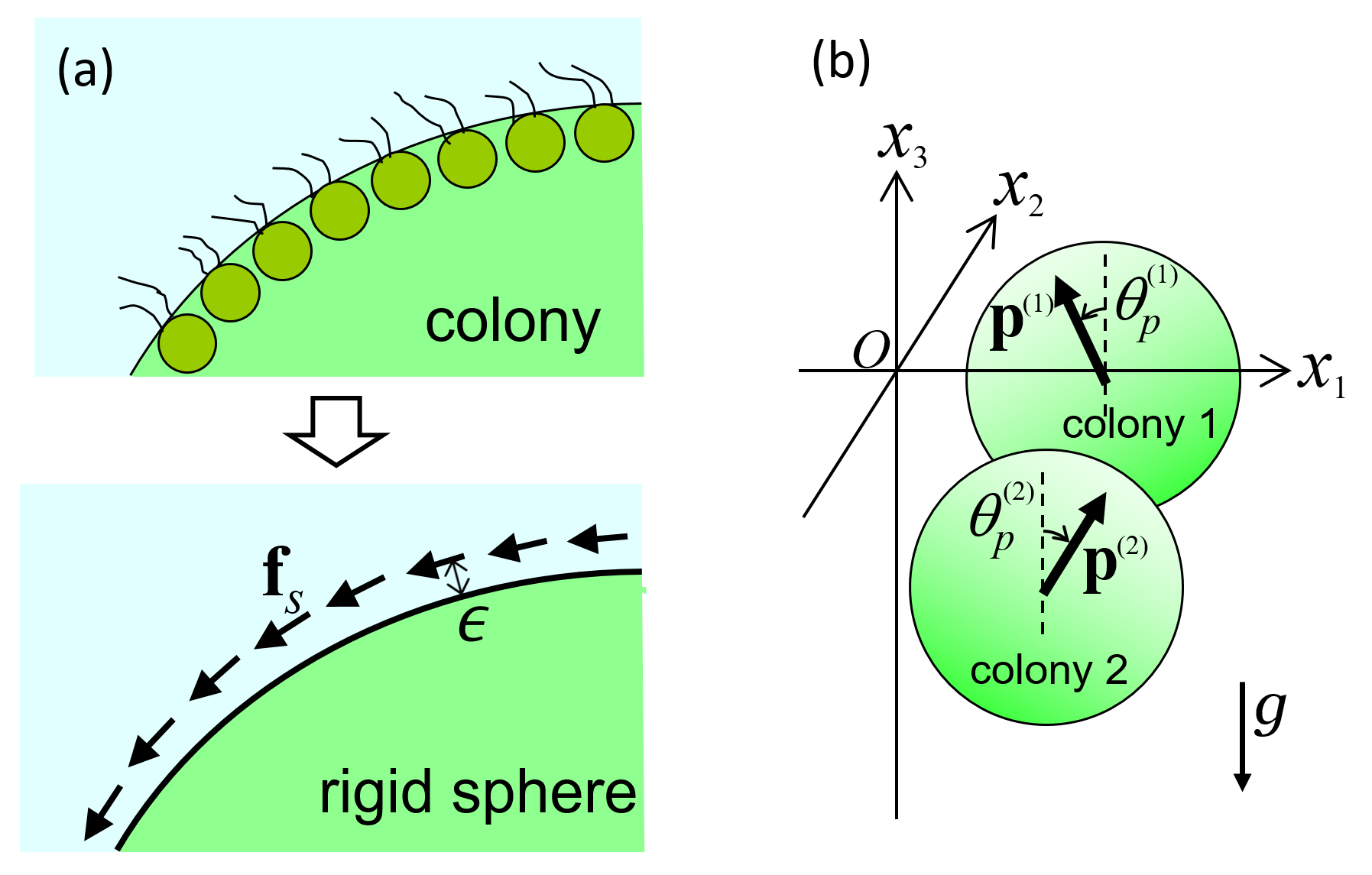}
\caption{Fluid mechanical model of \emph{Volvox}. (a) The colony is modeled as a rigid sphere, and forces generated by flagella are expressed by a shell of shear stress ${\bf f}_s$ at the distance e above the spherical surface. (b) Cartesian coordinate system used in the study, in which the gravity ${\bf g}$ acts in the ${\bf e}_3$ direction. A plane wall exists at $e_3 = 0$. The orientation vector of colony $m$ is ${\bf p}^{(m)}$ that has the angle $\theta_p^{(m)}$ from the ${\bf e}_3$ axis.
}
\label{fig4}
\end{center}
\end{figure}

\clearpage

\begin{figure}
\begin{center}
~~

\vspace{1cm}
\includegraphics[scale=0.4]{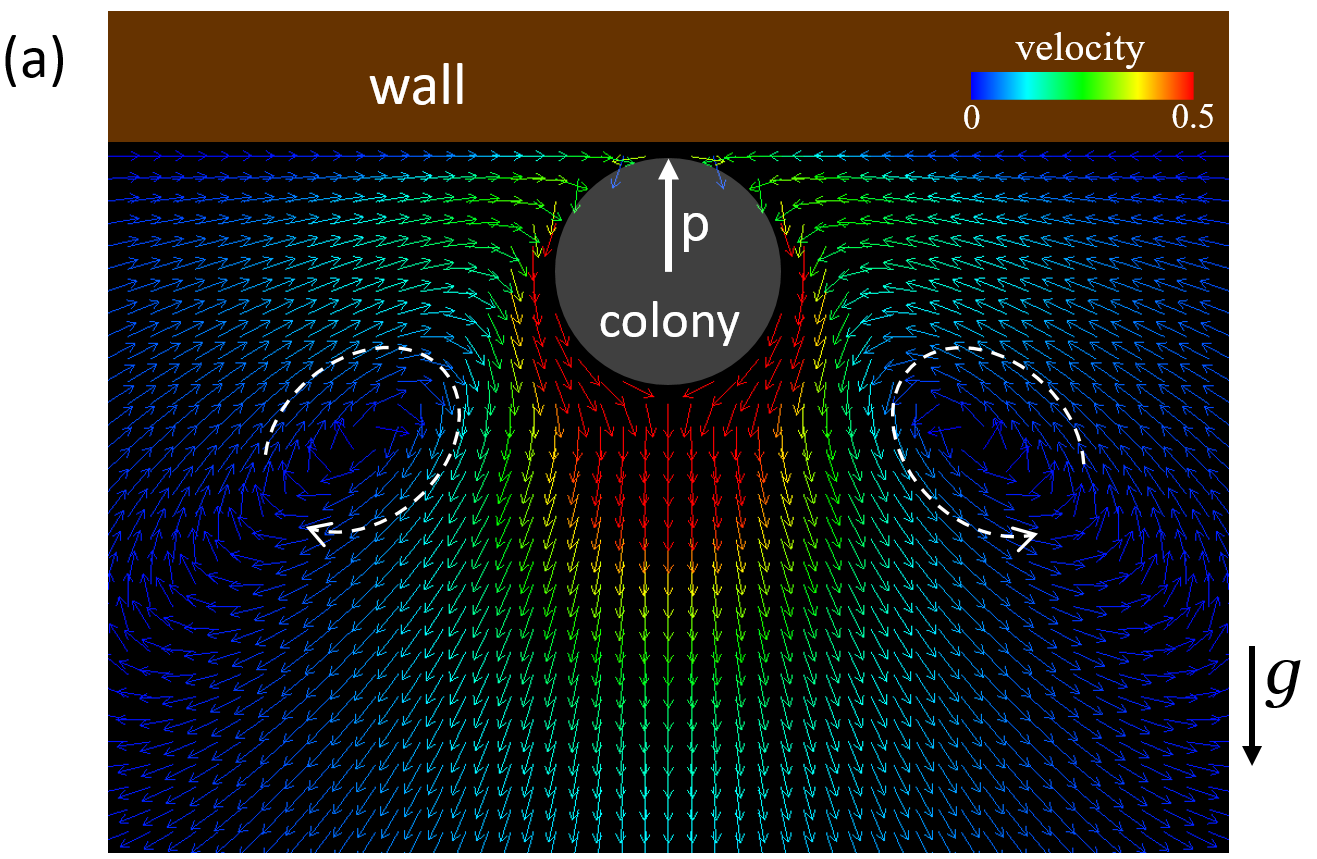}

\vspace{0.5cm}
\includegraphics[scale=0.4]{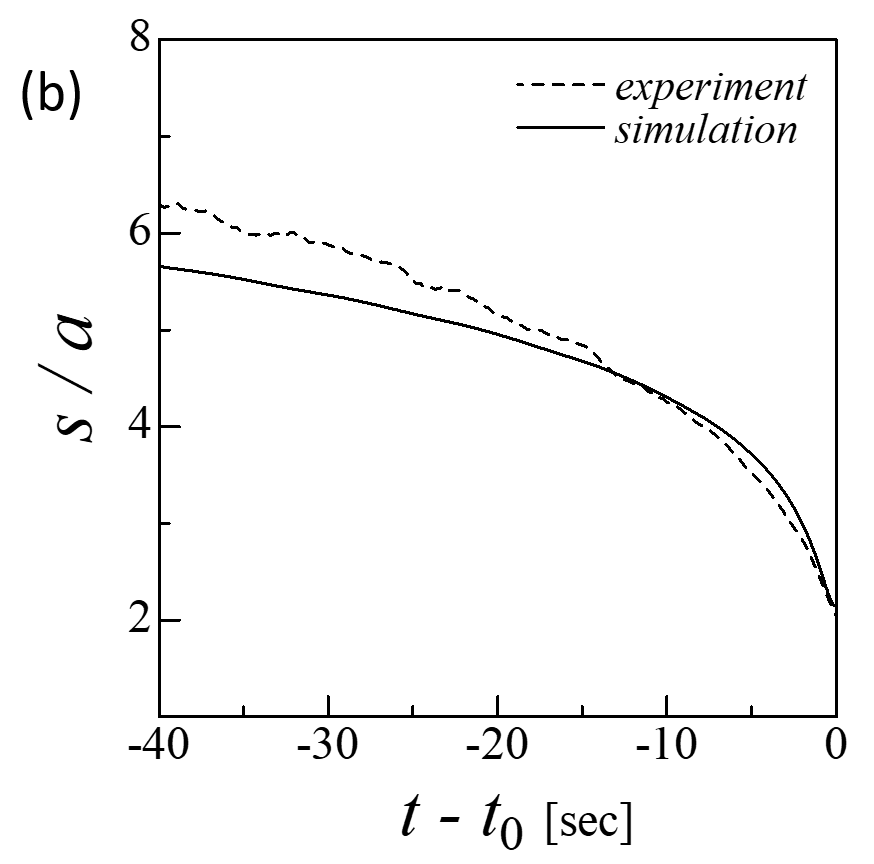}
\caption{A hovering colony beneath a top wall ($G_{bh} = 25$ and $F_g = 3 \pi$). (a) Velocity vectors around a stably hovering colony beneath a top wall. The colony is directed vertically upwards. White broken arrows schematically show the vortex structure. (b) Time change of center to center distance s between two colonies, where $t_0$ is the time of collision. The broken line indicates experimental result \cite{Drescher:2009vn}, and the solid line indicates our simulation result. The simulation result is dimensionlized by assuming that the colony swims one body length per second in the absence of gravity.
}
\label{fig5}
\end{center}
\end{figure}

\clearpage

\begin{figure}
\begin{center}
~~

\vspace{1cm}
\includegraphics[scale=0.22]{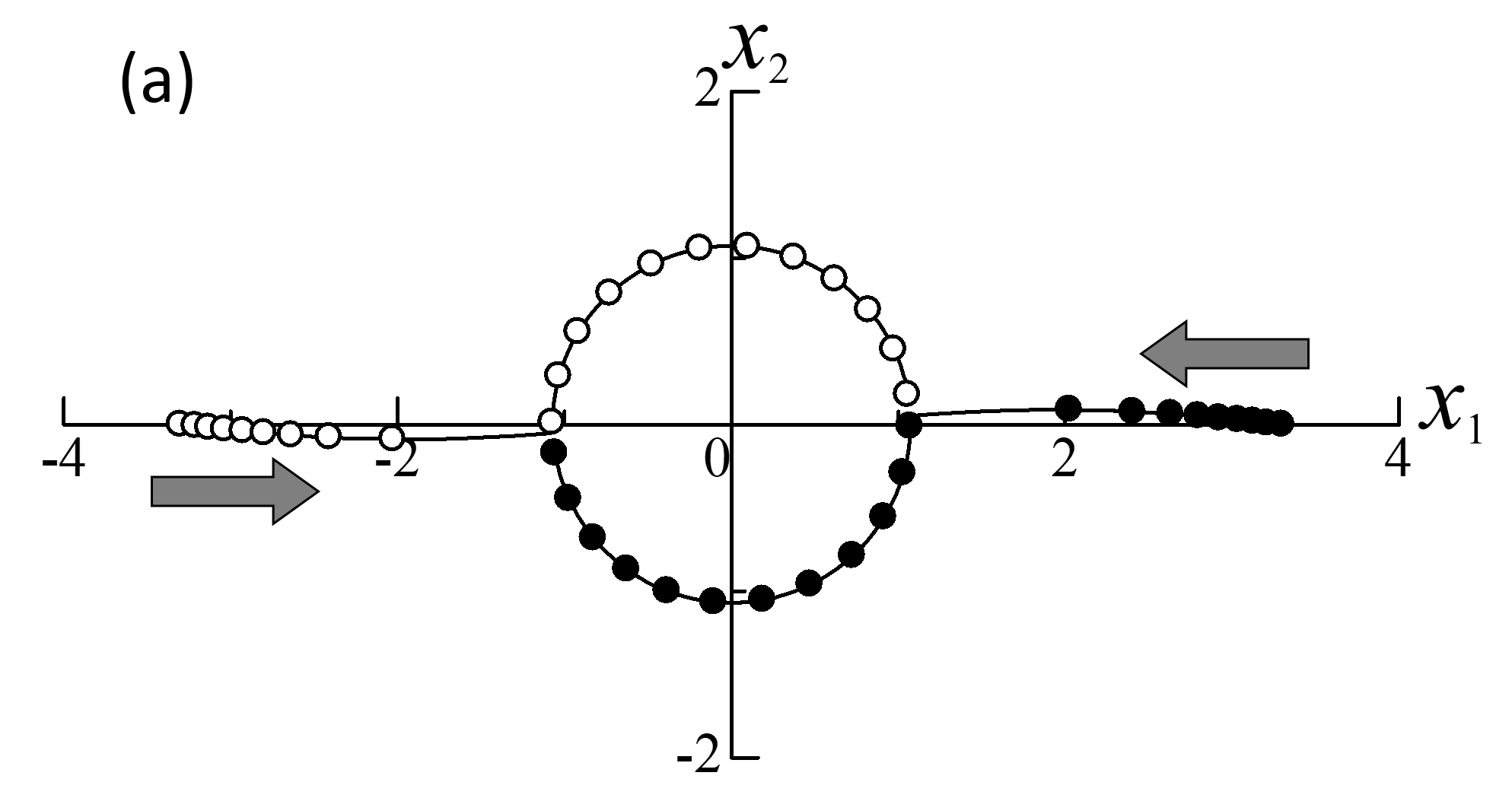}

\vspace{0.5cm}
\includegraphics[scale=0.25]{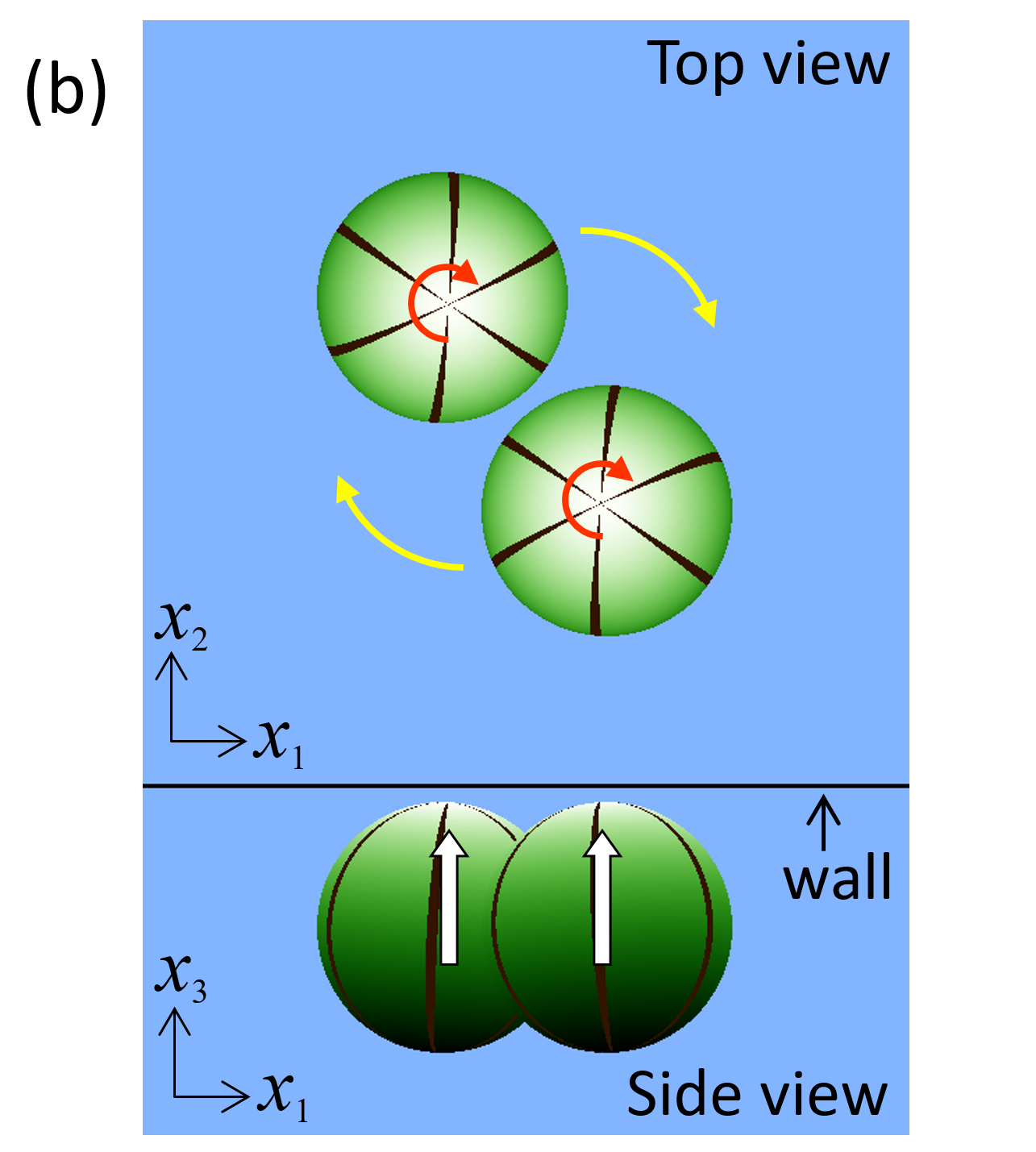}
\caption{Waltzing motion of two colonies ($G_{bh} = 25$ and $F_g = 3 \pi$). (a) Trajectories of two colonies. White or black circles indicate the centre positions of each colony, which are plotted with the time interval of $20a/U$. The colonies attracted each other and finally displayed waltzing motions. (b) Sample image of waltzing colonies, where two colonies are trapped just below the top wall and orbit around each other. Red and yellow arrows schematically show spin and orbit motions, respectively. (See supplementary Movie 3)
}
\label{fig6}
\end{center}
\end{figure}

\clearpage

\begin{figure}
\begin{center}
~~

\vspace{0.2cm}
\includegraphics[scale=0.2]{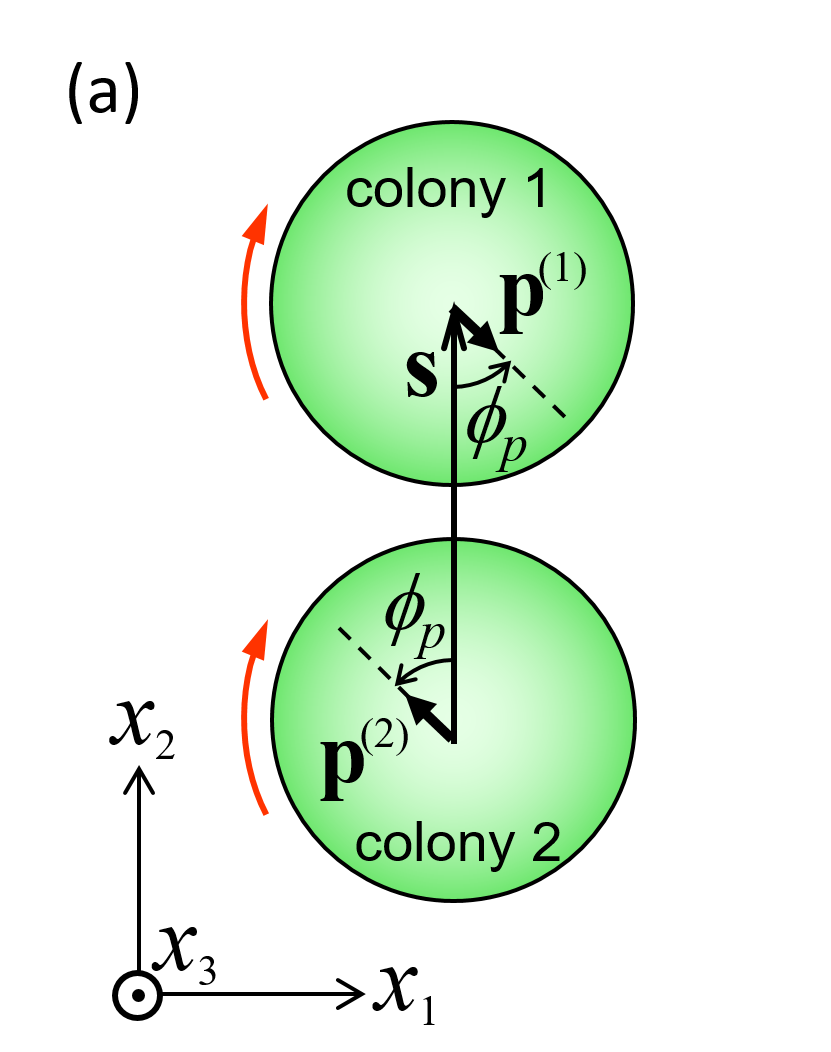}

\vspace{0.2cm}
\includegraphics[scale=0.35]{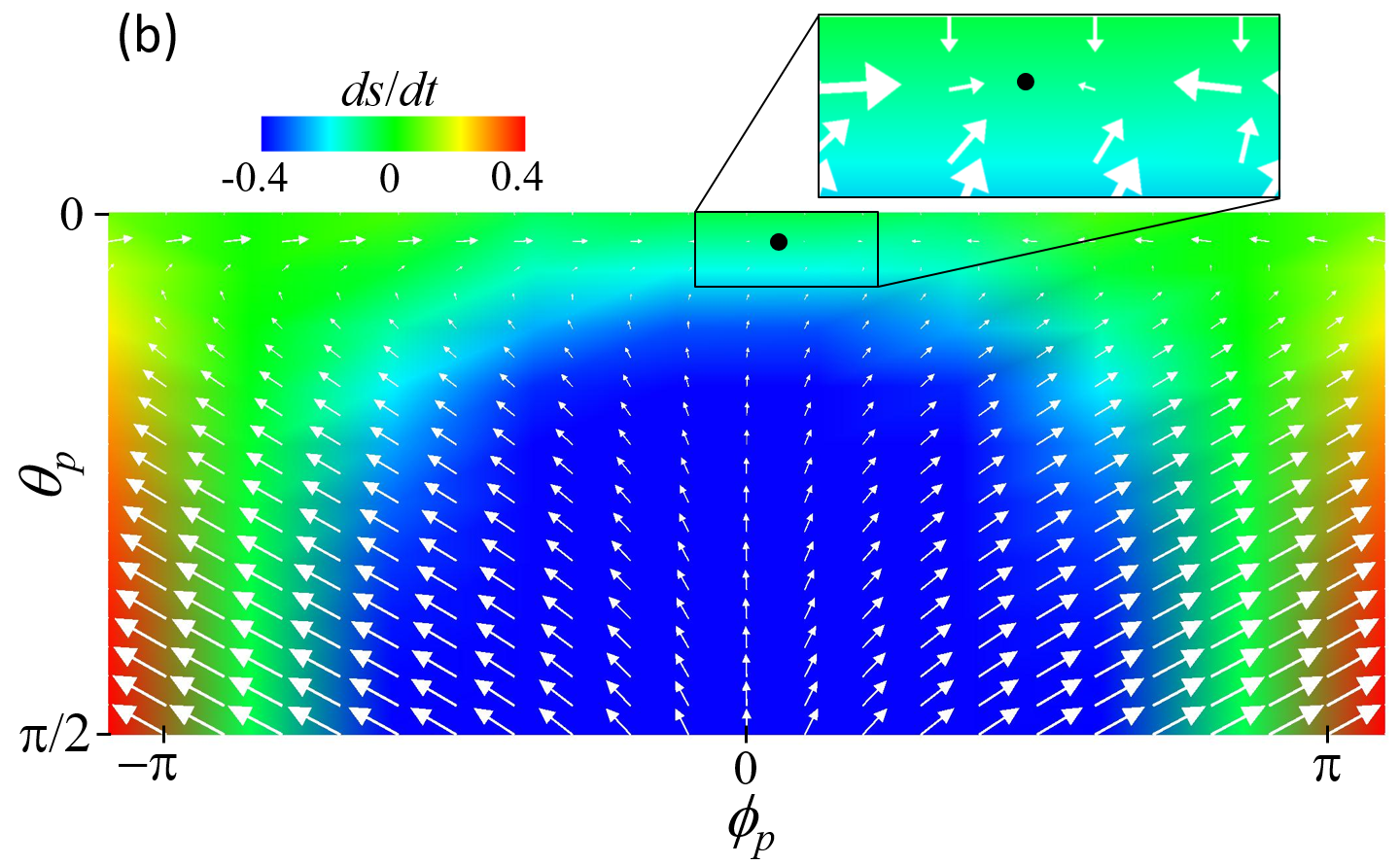}

\includegraphics[scale=0.35]{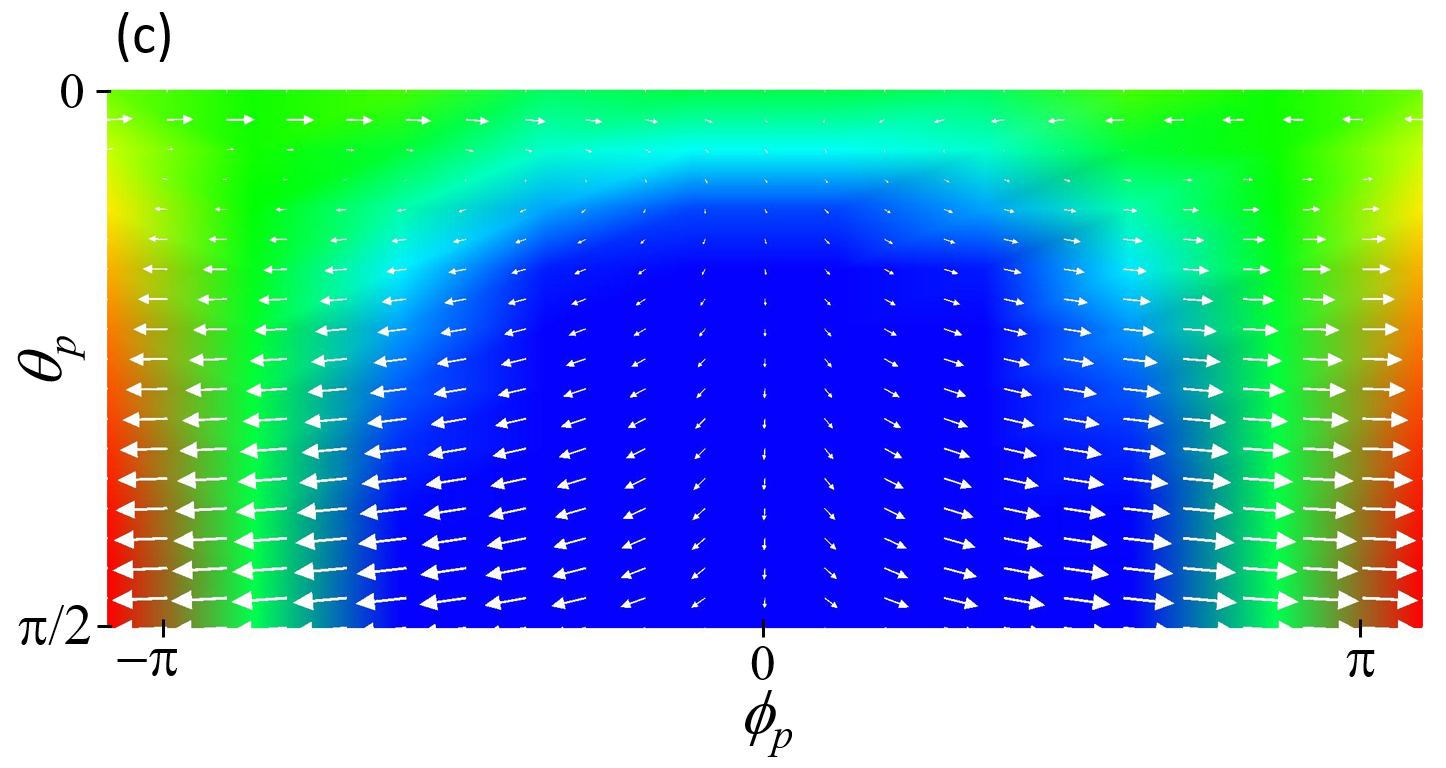}
\caption{Stability of waltzing motion. White vectors indicate the angular velocity in spherical cordinate $\theta_p - \phi_p$. Colours indicate the separation velocity of two colonies. (a) Definition of ${\bf s}$ and $\phi_p$. (b) Stability in the case of $G_{bh} = 25$ ($F_g = 3 \pi$). Stable waltzing motion is observed. Stable orientation ($\theta_p = 0.075, \phi_p = 0.092$) is shown by a black circle. Inset is the magnified image of the black rectangle. (c) Stability in the case of $G_{bh} = 5$ ($F_g = 3 \pi$). Waltzing motion is unstable.
}
\label{fig7}
\end{center}
\end{figure}

\clearpage

\begin{figure}
\begin{center}
~~

\vspace{5cm}
\includegraphics[scale=0.4]{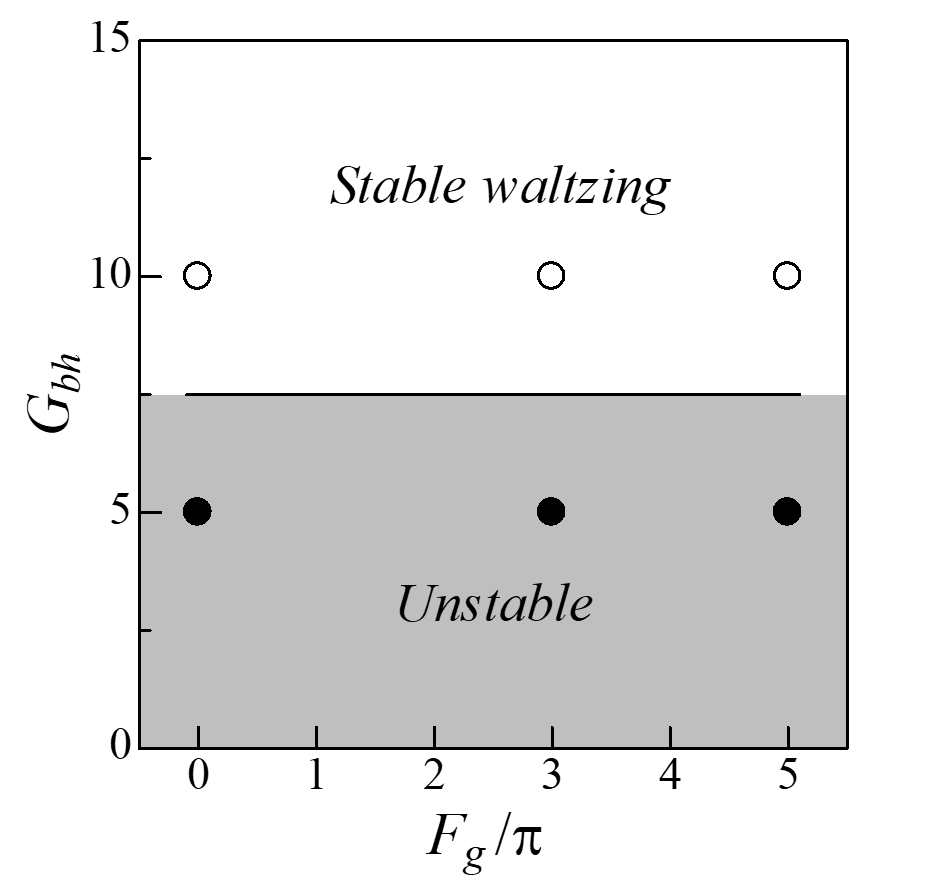}
\caption{Phase diagram on the stability of waltzing motion in $G_{bh} - F_g$ space. Circles indicate the simulation cases. The waltzing is unstable in the bottom grey region, while stable in the top white region
}
\label{fig8}
\end{center}
\end{figure}

\clearpage

\begin{figure}
\begin{center}
~~

\vspace{3cm}
\includegraphics[scale=0.3]{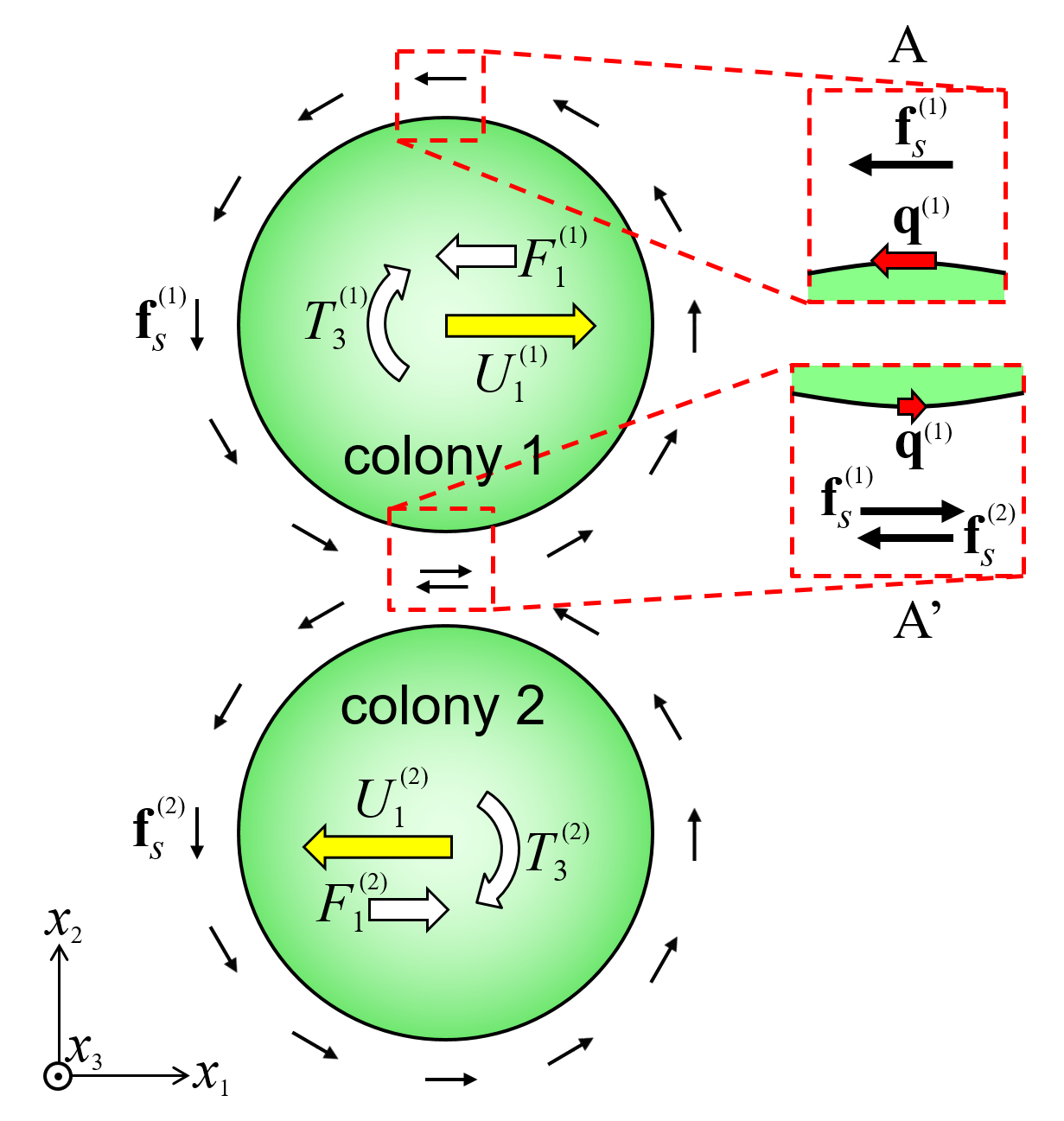}
\caption{Schematics of forces and torques exerted on two colonies fixed in space. ${\bf f}_s^{(m)}$ is the shear stress of colony $m$, and ${\bf q}_m$ is the traction generated on the surface of colony $m$. $F_1^{(m)}$ and $T_3^{(m)}$ are the $e_1$ component of the total force and the $e_3$ component of the total torque exerted on colony $m$, respectively. Magnified views of regions $A$ and $A'$ are indicated by the red broken lines.
}
\label{fig9}
\end{center}
\end{figure}

\clearpage

\begin{figure}
\begin{center}
~~

\vspace{3cm}
\includegraphics[scale=0.5]{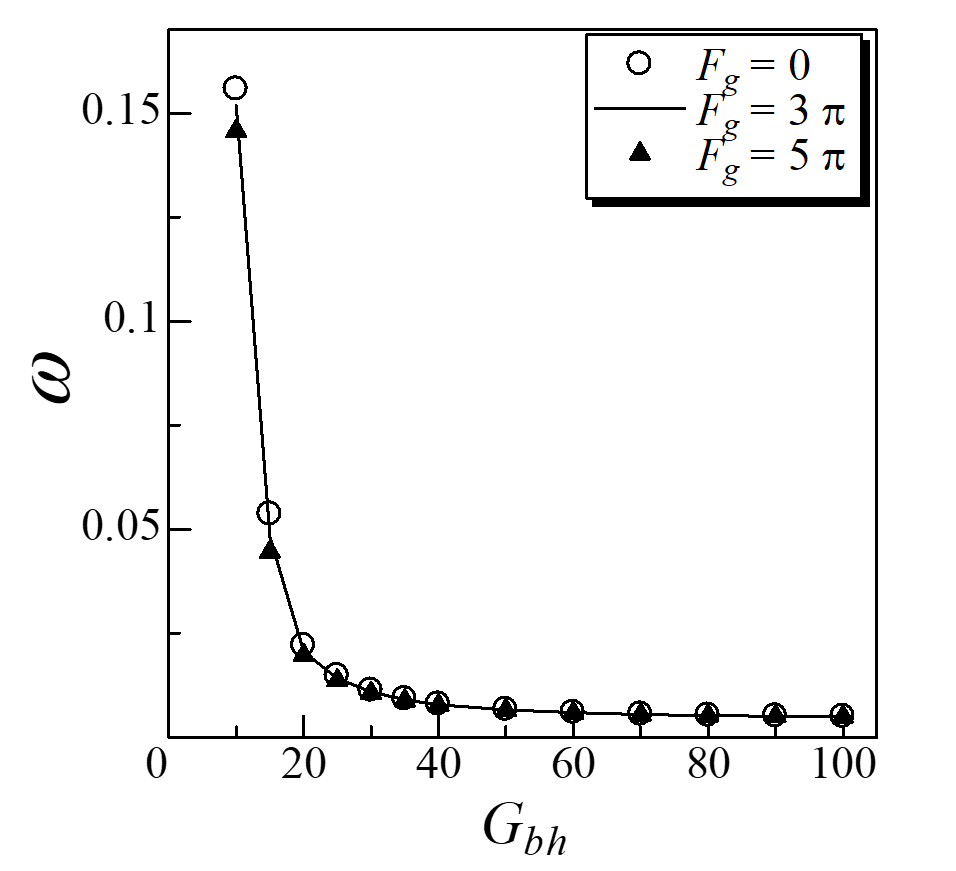}
\caption{Effect of $G_{bh}$ on the angular velocity of orbiting for various $F_g$ values
}
\label{fig10}
\end{center}
\end{figure}

\clearpage

\begin{figure}
\begin{center}
~~

\vspace{2cm}
\includegraphics[scale=0.23]{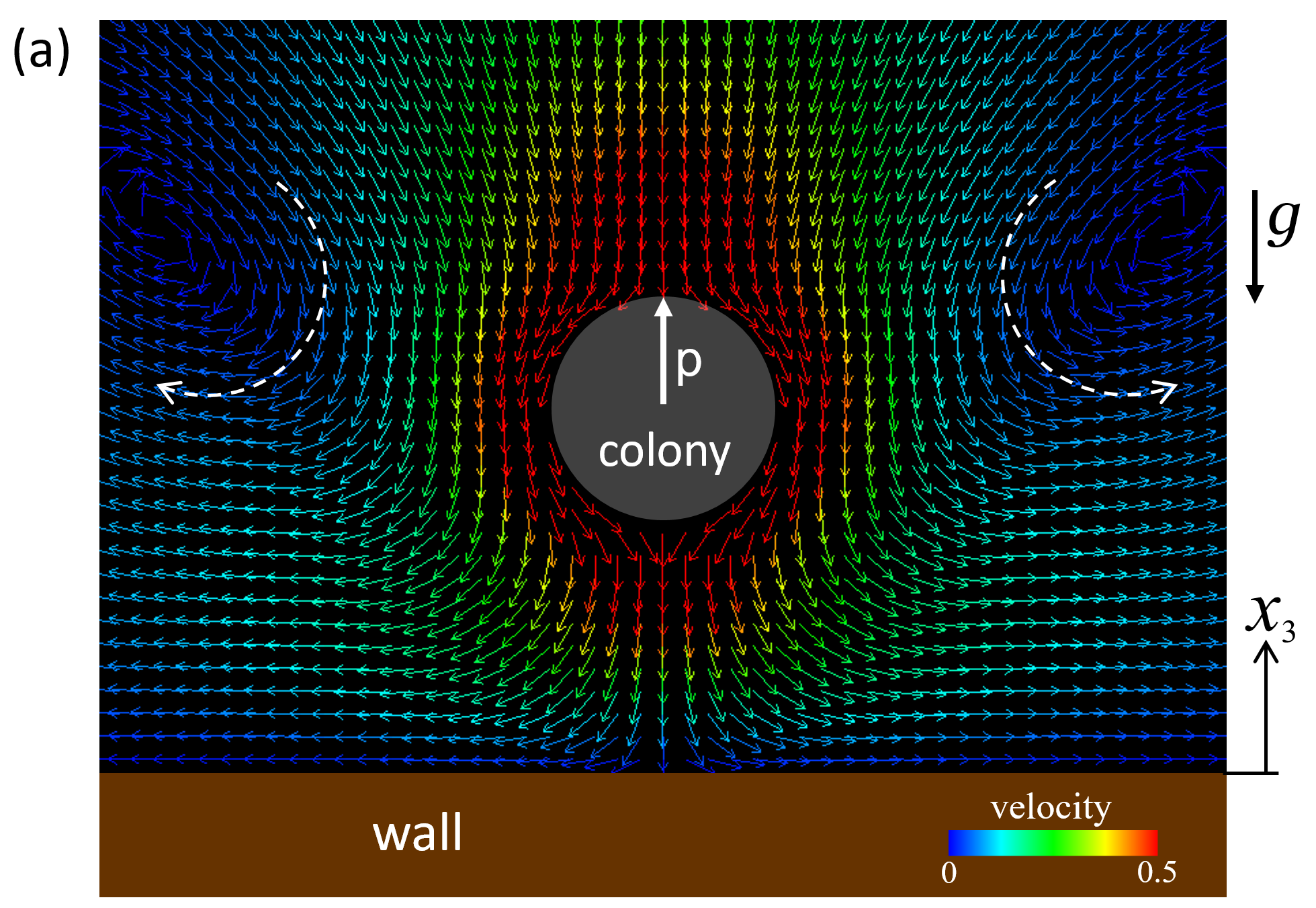}

\vspace{1cm}
\includegraphics[scale=0.42]{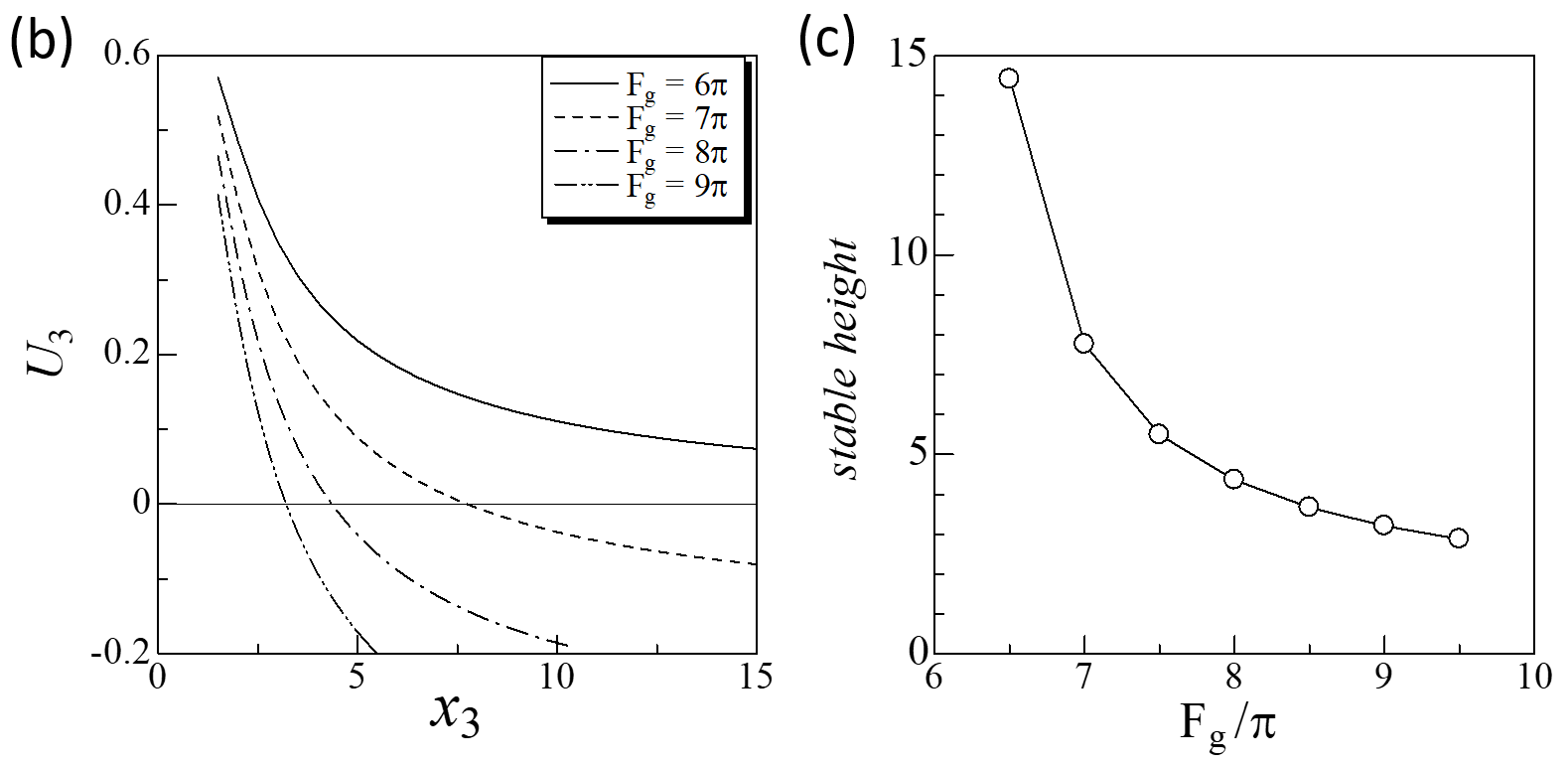}

\caption{Hovering of a colony near a bottom wall ($G_{bh}$ = 5). (a) Simulated velocity vectors around a stably hovering colony over a bottom wall ($F_g = 9 \pi$). The wall exists at $e_3$ = 0, and the $e_3$-axis is taken as shown in the figure. The colony is directed vertically upwards. White arrows schematically show the vortex structure. (b) Upward velocity of a single colony for various $F_g$ values. (c) Stable height of a hovering colony for various $F_g$ values.
}
\label{fig11}
\end{center}
\end{figure}

\clearpage

\begin{figure}
\begin{center}
~~

\vspace{2cm}
\includegraphics[scale=0.2]{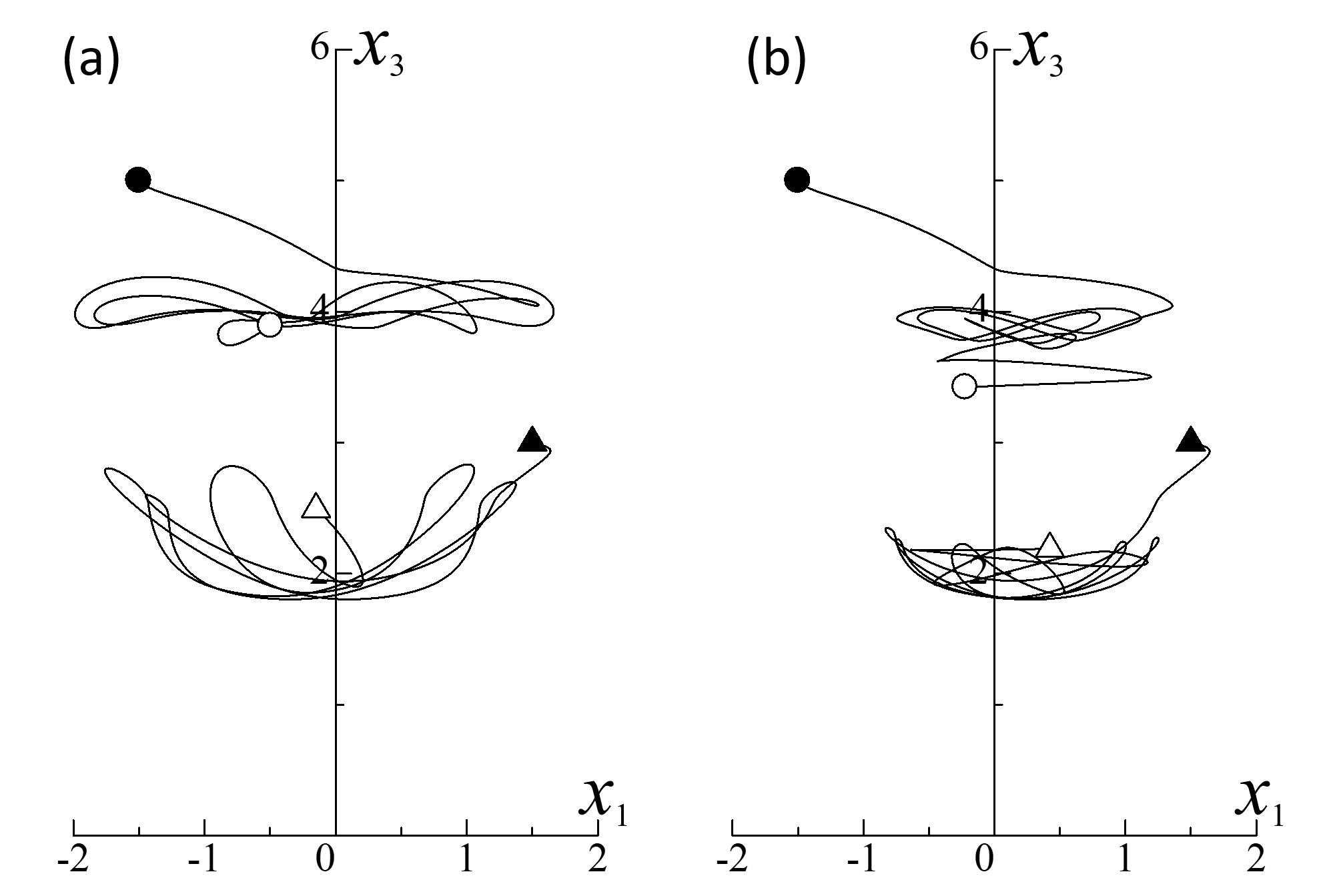}

\vspace{1cm}
\includegraphics[scale=0.2]{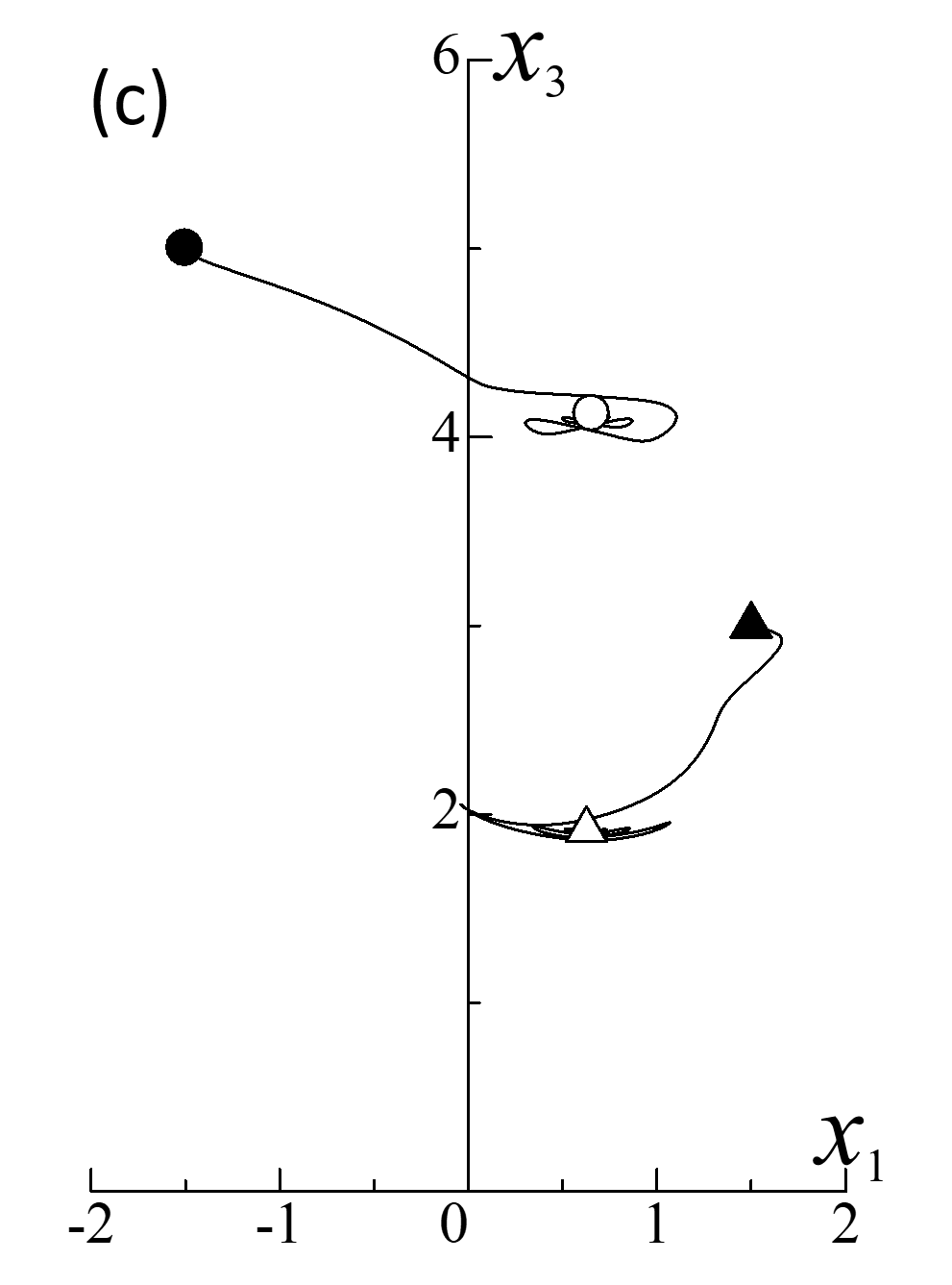}

\caption{Trajectories of two colonies near a bottom wall during $t = 0-100$. Trajectories of a colony with $F_g = 7.5 \pi$ start from the black circles and end at the white circles. Trajectories of a colony with $F_g = 9 \pi$ start from the black triangles and end at the white triangles. (a) Minuet motion with $G_{bh}$ = 2  (Supplementary Movie 4) (b) Minuet motion with $G_{bh}$ = 3  (Supplementary Movie 5) (c) Alignment of two colonies with $G_{bh}$ = 6  (Supplementary Movie 6)
}
\label{fig12}
\end{center}
\end{figure}

\clearpage

\begin{figure}
\begin{center}
~~

\vspace{5cm}
\includegraphics[scale=0.22]{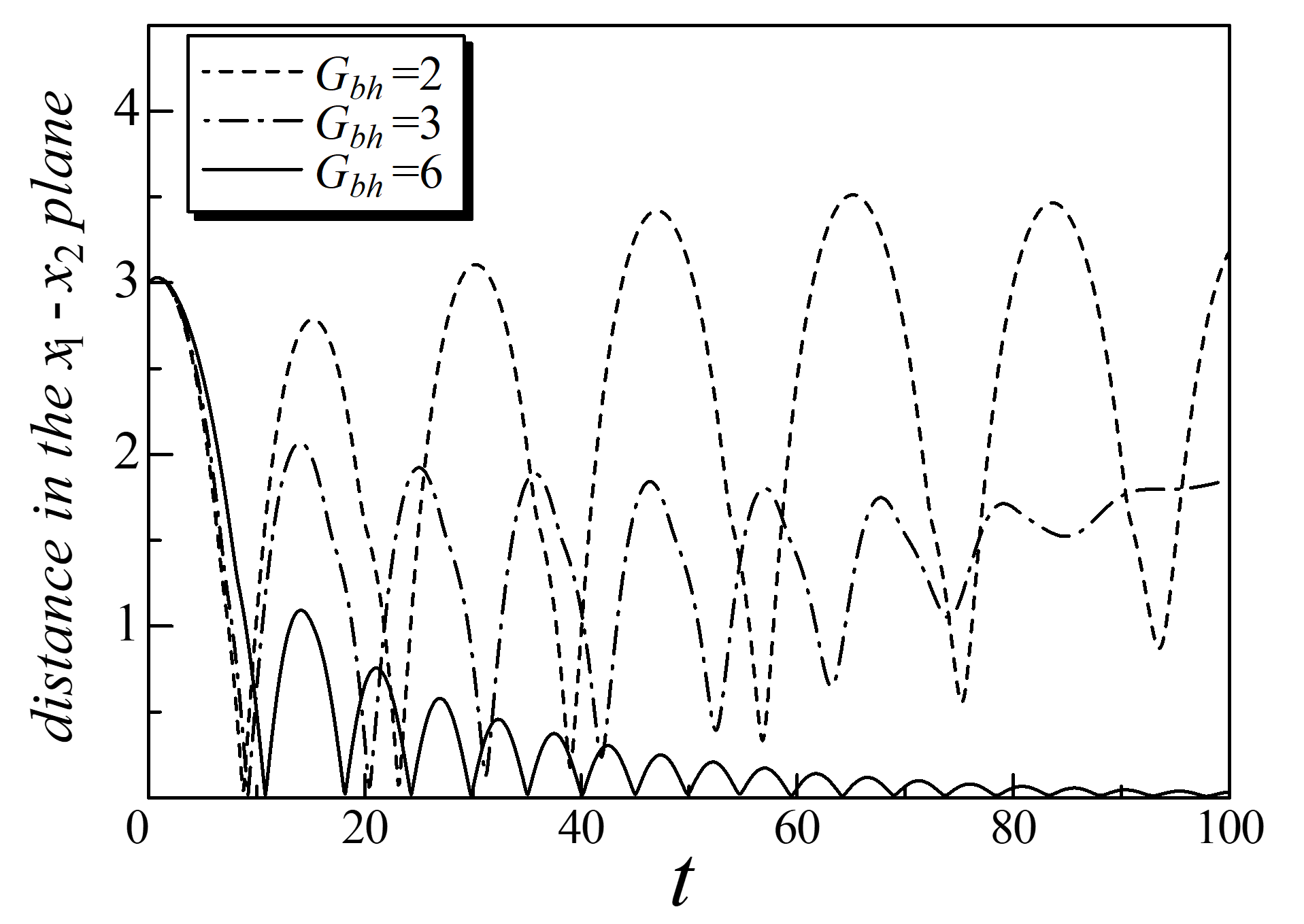}
\caption{Time course of the changing distance between two colonies with $F_g = 7.5 \pi$ and $9 \pi$ projected in the $e_1 - e_2$ plane. $G_{bh}$ is varied to 2, 3 and 6.
}
\label{fig13}
\end{center}
\end{figure}

\clearpage

\begin{figure}
\begin{center}
~~

\vspace{5cm}
\includegraphics[scale=0.5]{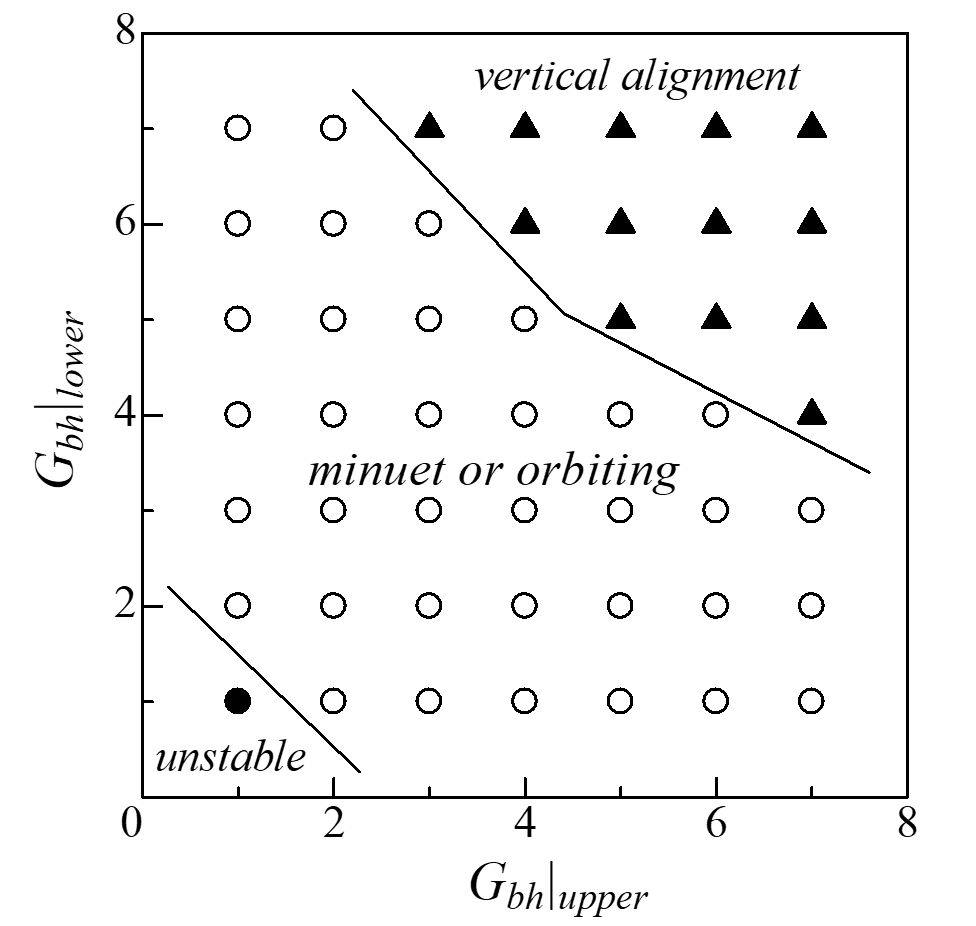}
\caption{Phase diagram of two \emph{Volvox} colonies interacting near a bottom wall ($F_g = 7.5 \pi$ and $9 \pi$). The black circle indicates 'unstable motion', in which the center to center distance between two colonies exceeds $10a$. The white circles indicate the 'minuet motion'. The black triangles indicate 'vertical alignment', in which the distance in the $e_1 - e_2$ plane is less than $0.3a$ during $t = 90-100$.
}
\label{fig14}
\end{center}
\end{figure}

\clearpage

\begin{figure}
\begin{center}
~~

\vspace{2cm}
\includegraphics[scale=0.22]{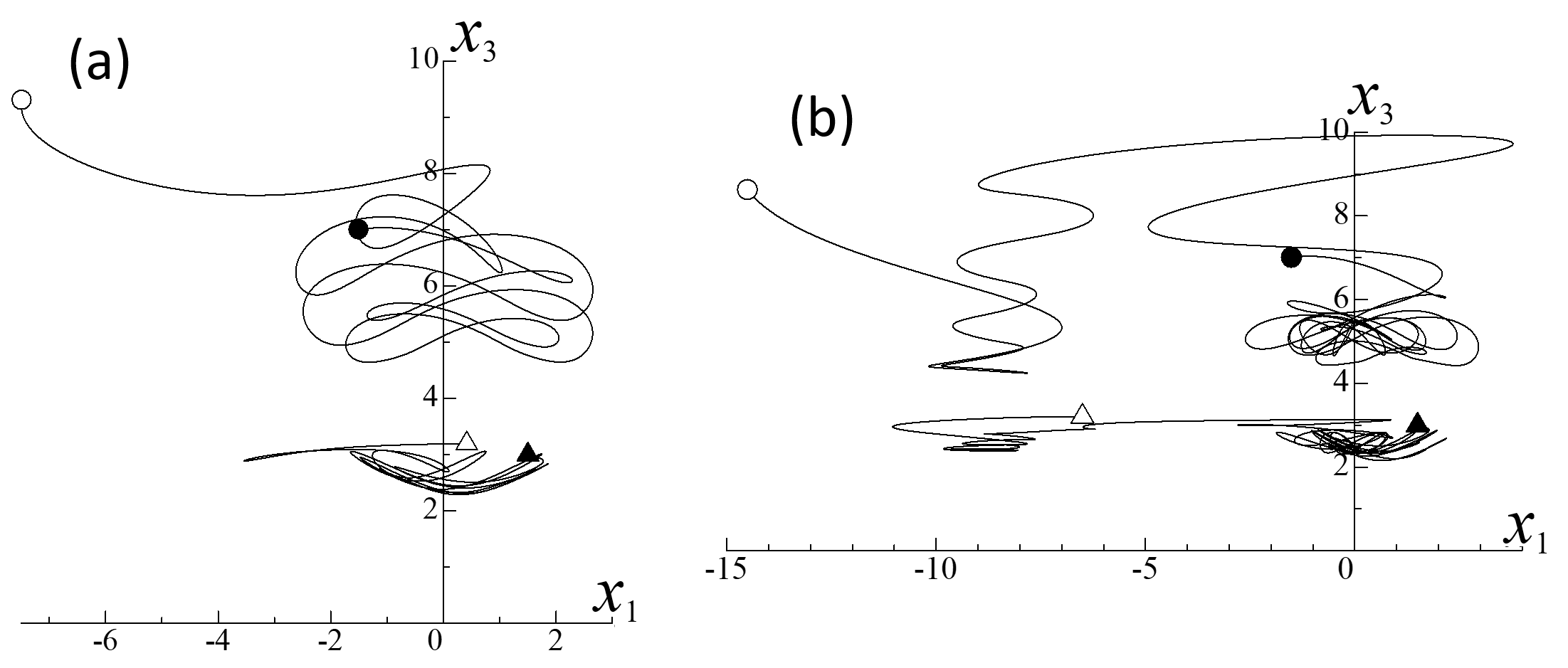}

\vspace{1cm}
\includegraphics[scale=0.22]{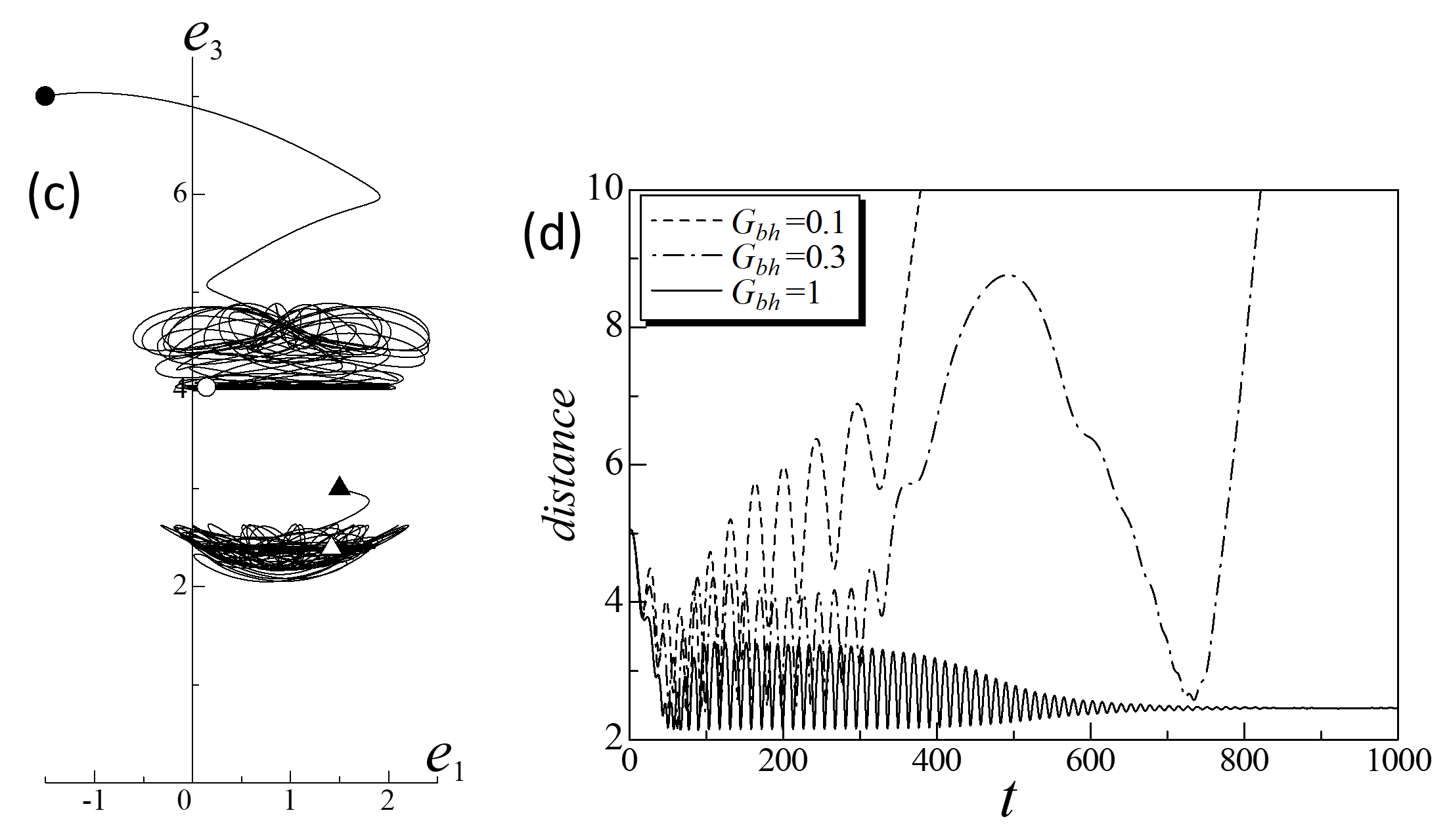}

\caption{Trajectories of two colonies near a bottom wall for time $t$ in the range $0-1000$ or until center-to-center distance exceeds $10a$. Trajectories of a colony with $F_g = 6.5 \pi$ start from the black circles and end at the white circles. Trajectories of a colony with $F_g = 9 \pi$ start from the black triangles and end at the white triangles. (a) Unstable far-field interaction with $G_{bh}$ = 0.1 (b) Unstable near-field interaction with $G_{bh}$ = 0.3 (c) Stable bound state with $G_{bh}$ = 1. Two colonies first show minuet motion, and then orbit around each other (d) Time change of the center-to-center distance of two colonies in (a), (b) and (c).
}
\label{fig15}
\end{center}
\end{figure}

\clearpage

\begin{figure}
\begin{center}
~~

\vspace{5cm}
\includegraphics[scale=0.4]{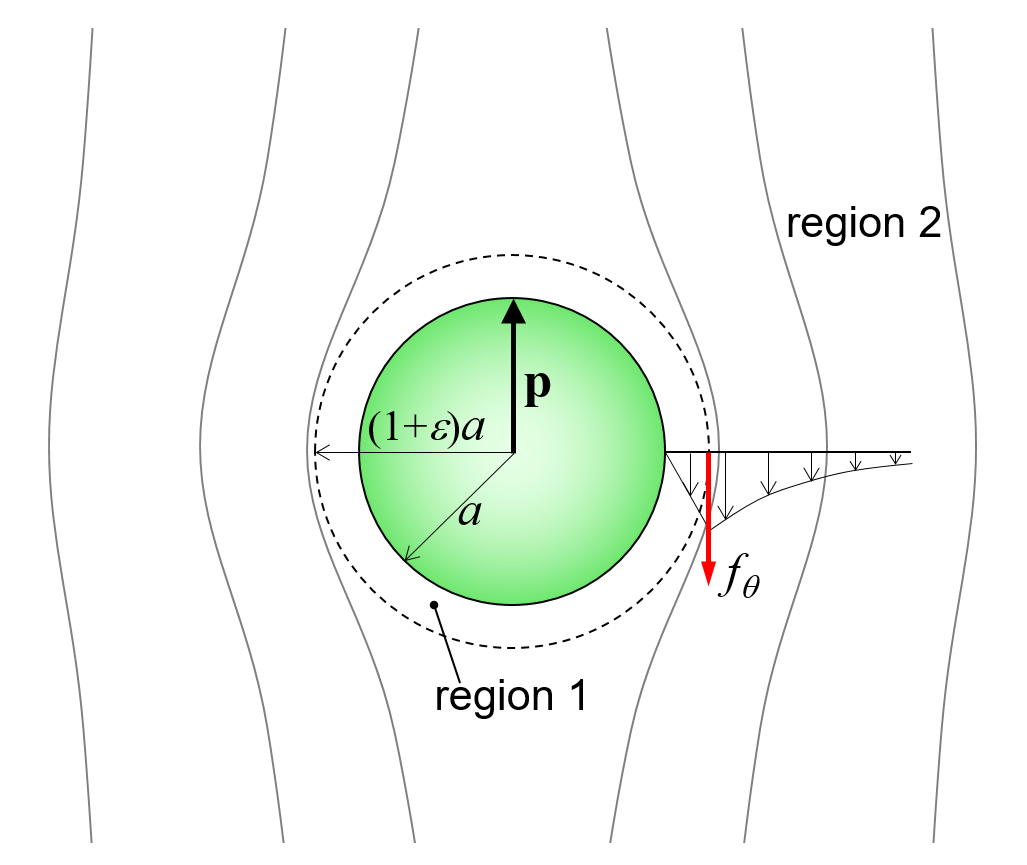}
\caption{
Schematic of the flow field around the 'shear-stress and no-slip' squirmer model. There is a no-slip spherical boundary at $r=a$, and uniform tangential stresses $f_\theta$ and $f_\phi$ are applied to the fluid at radius $(1+\varepsilon)a$. Region 1 is defined as $a < r < (1+\varepsilon)a$, whereas region 2 is defined as $(1+\varepsilon)a < r$.
}
\label{fig16}
\end{center}
\end{figure}

\clearpage

\begin{flushleft}

~~
\vspace{10mm}

{\Large Captions to movies}

\vspace{5mm}

Movie 1: Waltzing of \emph{Volvox carteri} observed by Drescher et al. (2009).
Three colonies are dancing near a top glass wall,
which is seen from the top.

\vspace{5mm}

Movie 2: Minuet of \emph{V. carteri} observed by Drescher et al. (2009).
Colonies are dancing near a bottom wall, which is seen from
the side.

\vspace{5mm}

Movie 3: Waltzing motion of two colonies with $G_{bh} = 25$ and $F_g = 3\pi$),
shown in Fig. 6.

\vspace{5mm}

Movie 4: Minuet motion of two colonies with $G_{bh} = 2$, shown in Fig. 12a.

\vspace{5mm}

Movie 5: Minuet motion of two colonies with $G_{bh} = 3$, shown in Fig. 12b.

\vspace{5mm}

Movie 6: Minuet motion of two colonies with $G_{bh} = 6$, shown in Fig. 12c.

\end{flushleft}

\end{document}